\documentclass[12pt, a4paper]{article}
\usepackage[a4paper, left=3cm,right=2cm]{geometry}
\usepackage{hyperref}
\usepackage{amsfonts} 
\usepackage{amsmath}
\usepackage{amssymb}
\usepackage{setspace}
\usepackage{color}

\usepackage[utf8,applemac]{inputenc}
\usepackage{tensor}
\usepackage{cite}
\usepackage{tikz}
\usepackage{graphicx,subfigure}% Include figure files
\usepackage[font=small, font+=it, format=hang, margin={1cm, 1cm}]{caption}
\graphicspath{{figure/}}

\usepackage{dcolumn}% Align table columns on decimal point
\usepackage{bm}% bold math

\numberwithin{equation}{section}

\newcommand{\bea}{\begin{eqnarray}}
\newcommand{\eea}{\end{eqnarray}}
\newcommand{\be}{\begin{equation}}
\newcommand{\ee}{\end{equation}}
\def\nn{\nonumber}

\def\p{\partial}
\def\eps{\epsilon}

\newcommand{\cR}{\mathcal{R}}

\newcommand{\cM}{\mathcal{M}}
\newcommand{\cW}{\mathcal{W}}

\newcommand{\cK}{\mathcal{K}}

\renewcommand{\d}{\textrm{d}}

\begin{document}

%\maketitle
%\abstract{[...]}

\setcounter{tocdepth}{2}

\begin{titlepage}

\begin{flushright}\vspace{-3cm}
{\small
%{\tt arXiv:yymm.nnnn} \\
\today }\end{flushright}
\vspace{0.5cm}

\begin{center}

{{ \LARGE{\bf{Spin and Quadrupole Couplings for\\
 High Spin Equatorial \vspace{6pt}\\
 Intermediate Mass-ratio Coalescences }}}}
\vspace{5mm}

\bigskip
\bigskip

\centerline{\large{\bf{Bin Chen$^{\star\ast\bullet}$, Geoffrey Comp\`{e}re$^\diamondsuit$, Yan Liu$^\diamondsuit$, Jiang Long$^\ddagger$}}} \vspace{5pt}
\centerline{\large{\bf{and  Xuao Zhang$^\star$\footnote{emails: bchen01@pku.edu.cn, gcompere@ulb.ac.be, yliu6@ulb.ac.be, jiang.long@apctp.org, stringwaves@pku.edu.cn}}}}

\vspace{2mm}
\normalsize
\bigskip\medskip
\textit{{}$^{\star}$Department of Physics and State Key Laboratory of Nuclear Physics and Technology,\\Peking University, 5 Yiheyuan Rd, Beijing 100871, P.~R.~China}\\
\vspace{2mm}
\textit{{}$^{\ast}$Collaborative Innovation Center of Quantum Matter, 5 Yiheyuan Rd, Beijing 100871, P.~R.~China}\\
\textit{{}$^{\bullet}$Center for High Energy Physics, Peking University, 5 Yiheyuan Rd, Beijing 100871, P.~R.~China}\\\vspace{2mm}

\textit{{}$^\diamondsuit$ Universit\'{e} Libre de Bruxelles, Gravitational Wave Centre, \\
 International Solvay Institutes, CP 231, B-1050 Brussels, Belgium}\\\vspace{2mm}
\textit{{}$^\ddagger$ Asia Pacific Center for Theoretical Physics, Pohang 37673, Korea}

\bigskip

\vspace{15mm}

\begin{abstract}
\noindent
{Intermediate mass-ratio coalescences are potential signals of ground-based and space-based gravitational observatories. Accurate modeling of their waveforms within general relativity can be achieved within black hole perturbation theory including self-force and finite size effects. In this paper, we present analytic results to the Teukolsky perturbation of equatorial orbits in the near-horizon region of  an extremely high spin black hole including spin coupling and finite size effects at leading order in the high spin limit while neglecting the self-force. We detail the critical behavior occuring close to the smallest specific angular momentum, and we discuss features of spin and quadrupole couplings. 
}

\end{abstract}

%\pacs{04.65.+e,04.70.-s,11.30.-j,12.10.-g}
%PACS: 04.20.-q, 04.20.Ha, 11.25.Tq, 11.30.Cp

\end{center}
%%%%%%%%%%%%%%%%%%%%%%%%%%%%%%%%%%%%%%%%%%%%%%%%%%%%%%%%%%%%%%%%%%%%%%%%%%%%%%%%%%%%%%%%

\end{titlepage}

\tableofcontents

\section{Introduction}

Intermediate mass ratio coalescences (IMRACs) are binary systems of mass ratio $q \sim 10^{-2}-10^{-4}$, which involve a black hole with mass in the range $100-10^5 M_\odot$, a so-called intermediate-mass black hole (IMBH). There are two types of IMRACs: Type I consists of a stellar mass compact object inspiraling and merging into an IMBH, while Type II consists of an IMBH inspiraling and merging into a supermassive black hole (SMBH)\footnote{In the literature, such systems are alternatively termed intermediate mass-ratio inspirals (IMRIs). For IMRACs of Type I, merger and ringdown contribute significantly to the gravitational-wave signal of ground-based detectors \cite{Smith:2013mfa} while for IMRACs of Type II, merger and ringdown contribute significantly to the signal of space-based detectors \cite{Miller:2004va}. We therefore adopt the terminology IMRAC instead of IMRI.}.

Type I IMRACs can be formed in dense stellar systems such as galactic nuclei and globular clusters, as shown by stellar-dynamics simulation of globular clusters\cite{Konstantinidis:2011ie,Leigh:2014oda,Haster:2016ewz,MacLeod:2015bpa} and the parabolic capture mechanism\cite{1989ApJ...343..725Q}. In this case, the central black holes are IMBHs with masses in the range from a few hundred to a few thousand solar masses, and the orbiting compact objects have stellar mass.  Depending on their orbital parameters, IMRACs in clusters can be detectable\cite{Amaro-Seoane:2018gbb} not only by the space-based gravitational wave (GW) observatories, including planned LISA \cite{Audley:2017drz,AmaroSeoane:2007aw} and prospective  Tian Qin \cite{Luo:2015ght} or Taiji \cite{Guo:2018npi}, but also by the ground-based observatories LIGO/Virgo \cite{Brown:2006pj,Mandel:2007hi} or the Einstein Telescope (ET) \cite{Huerta:2010un,Huerta:2010tp}. This also leads to the possibility of LISA first detecting signals and warning a year in advance the ground based detectors \cite{Amaro-Seoane:2018gbb}. For intermediate mass-ratio coalescences of Type I, merger and ringdown contribute significantly to
the gravitational-wave signal of ground-based detectors \cite{Smith:2013mfa}, which particularly motivates our present analysis. Type II IMRACs are a potentially very loud source in the LISA band though sparse in the number of events. The GW signal may have a large signal-to-noise ratio (SNR)  and can be detectable during the end of the inspiral without the need for matched filtering \cite{Miller:2004va}. 

IMBHs could be primordial, generated in the early  Universe, or they may form in the center of dense globular clusters through runaway stellar collision. They are thought to be the seeds from which SMBHs grow. The study of IMBHs may shed light on various astrophysical problems in galaxy formation and growth, black hole accretion and reionisation \cite{Miller:2003sc}.  The observational evidence for the existence of IMBHs has accumulated in the past decade thanks to X-ray  and optical observations \cite{Mezcua:2017npy}. A single observation of GW from a IMRAC would be a direct detection of an IMBH and it would provide with an accurate mass measurement.  As the population of IMBHs is poorly understood, the event rates for both types of IMRACs are not well constrained. In \cite{Gair:2010dx}, it was estimated that the type I IMRAC event rate for ET could be as high as few hundred per year. For the type II IMRAC event rate for LISA, the preliminary estimations range from a few events per year \cite{Miller:2004va} to a hundred events per year \cite{PortegiesZwart:2005zp,Gair:2010dx}. The typical timescale of the detectable signal at ET for a Type I IMRAC is a few seconds, while for a Type II IMRAC at LISA is a few minutes. This is in sharp contrast with the years of signal of the extreme mass ratio inspirals (EMRIs) at LISA.

The gravitational waveform emitted from the late inspiral to the merger of an IMRAC encodes detailed information about the strong field region near the central black hole. Precision tests of general relativity and the potential discovery of new fundamental physics requires an extremely accurate waveform modeling of the entire detectable signal \cite{Barack:2018yly}, which for IMRACs includes the merger. For comparable mass binary systems, this is achieved within the effective one-body formalism \cite{Buonanno:1998gg,Damour:2001tu} (see \cite{Damour:2011xga} for a review), informed by high order post-Newtonian theory (see \cite{Blanchet:2013haa} for a review), post-Minkowskian expansions \cite{Damour:2016gwp} and complete numerical $3+1$ simulations \cite{Jani:2016wkt}. The post-Newtonian expansion does not converge beyond the light-ring \cite{Akcay:2012ea} and therefore requires non-perturbative techniques in the velocity expansion to model  the transition of the inspiral to the merger \cite{Buonanno:2000ef}. The merger phase can be modelled with numerical simulations, which however become prohibitive in terms of computational power for a mass ratio higher than ten because of the separation of scales.  In the opposite extreme mass ratio limit, waveforms can be derived in black hole perturbation theory including finite size effects and self-force corrections (see the reviews \cite{Poisson:2011nh,Harte:2014wya,Pound:2015tma,Barack:2018yvs}). The transition from the inspiral to the merger has been modeled so far only taking into account the orbit-averaged, dissipative piece of the self-force \cite{Ori:2000zn,Kesden:2011ma,Compere:2019aa}. For IMRACs, a comprehensive modelization has not yet been performed as they lie somewhere between these two regimes. Perturbation theory has been shown to be a promissing approach to model IMRACs \cite{Tiec:2014lba}. Spin and finite size effects have been shown to be relevant for precision modeling \cite{Huerta:2011kt,Huerta:2011zi} and have been computed for binaries involving a central Schwarzschild black hole \cite{Warburton:2017sxk}, but such effects have not yet been systematically included within the black hole perturbation framework. 

The objective of this paper is to provide new accurate semi-analytical data in the modeling of the transition between the late inspiral and merger of binaries with a small or intermediate mass ratio by including spin and finite size effects. We will provide this semi-analytical data only in the case where the spin of central massive black hole is close to maximal and for a binary whose orbital plane is the equatorial plane (i.e. no inclination). As already emphasized, this modeling is most relevant for IMRACs since for EMRIs the merger phase is marginally observable. 

One key point in our study is that  the innermost stable circular orbit (ISCO) lies within the near-horizon region. As shown in \cite{Bardeen:1999px},  the ISCO asymptotically approaches the horizon location in Boyer-Linquist coordinates in the extremal limit. To describe the physics around the ISCO, one may introduce  a set of new coordinates in terms of  which the near-horizon geometry becomes the so-called NHEK geometry plus $O(\lambda^{1/3})$ corrections with $\lambda=\sqrt{1-J^2/M^4} \ll 1$. Given this fact, all results obtained in the near-horizon region around the ISCO are the leading results in the high spin expansion. The corrections go as $O(\lambda^{1/3})$, which is of the order of $40\%$ for Thorne's upper spin limit $J=0.998 M^2$ but negligible for extremely high spin. 

The spacetime around a high spin black hole can be viewed as a match between a ``very near-horizon region'', a ``near-horizon region'' and the exterior region diffeomorphic to extremal Kerr spacetime \cite{Bardeen:1972fi}. Both the ``very near-horizon region'' (near-NHEK) and the ``near-horizon region'' (NHEK) are diffeomorphic to each other and admit exact conformal $SL(2,R) \times U(1) \times PT$ symmetries that extend the $\mathbb R  \times U(1) \times PT$ symmetries of the Kerr geometry \cite{Bardeen:1999px,Amsel:2009ev,Bredberg:2009pv} (see \cite{Bredberg:2011hp,Compere:2012jk,Compere:2018aar} for reviews). The opening up of extended near-horizon regions at high redshift around high spin black holes gives rise to very specific strong gravity physics related to critical phenomena  \cite{Gralla:2016jfc,Compere:2017zkn,Gralla:2017lto,Compere:2017hsi,Gralla:2017ufe} and to specific conformal symmetry methods that can be used to derive exact semi-analytical GW waveforms \cite{Gralla:2016qfw,Porfyriadis:2014fja,Hadar:2014dpa,Hadar:2015xpa,Gralla:2015rpa,Hadar:2016vmk,Compere:2017hsi}. Therefore the waveforms computed using conformal symmetry methods are relevant from the transition between the inspiral and merger up to the final ringdown.

In this paper, we extend the status of conformal symmetry methods for GW \cite{Compere:2017hsi} to take spin and finite size effects into account within the Teukolsky formalism but using the probe approximation (no gravitational self-force). Self-force effects within the near-horizon geometry will be considered elsewhere \cite{Compere:2019bb}.We will use the Mathisson-Papapetrou-Dixon formalism \cite{Mathisson:1937zz,Papapetrou:1951pa,Dixon:1970zza,Dixon:1970zz,Dixon:1974aa} to model the extended black hole, neutron star or exotic compact object. The solution to the MPD equation that we will construct matches with the leading high spin limit of the prograde (with respect to the Kerr spin) circular orbit of \cite{Hinderer:2013uwa} after specializing to probe black holes.  Our semi-analytic results allow to derive the waveforms for an arbitrary plunging equatorial trajectory. This is an outcome of the conformal methods developed in \cite{Compere:2017hsi}. More precisely, our results are an extension of \cite{Compere:2017hsi,Gralla:2016qfw,Porfyriadis:2014fja,Hadar:2014dpa,Hadar:2015xpa,Gralla:2015rpa,Hadar:2016vmk}, including spin and finite size effects.  They provide with the waveforms of the final geodesic plunge of IMRACs at the leading order in the high spin limit with spin and finite size corrections.  However, these waveforms do not include the self-force effects which are of the same order in the mass-ratio as the spin effects. Our results on spin and finite size effects are the first step towards more realistic waveforms which include spin effects for the inspiral, transition and plunge.  

%as performed in \cite{Thorne:1984mz}.  

The rest of the paper is organized as follows. We first review of the Mathisson-Papapetrou-Dixon (MPD) formalism in Section 2 and review the asymptotically matched expansion scheme valid for a central high spin black hole in the context of black hole perturbation theory in Section \ref{HighSpinTeuk}. We solve the MPD equations for circular orbits in the near-horizon region of the Kerr black hole in Section \ref{sec:nearorbit}. We then derive the stress-tensor and solve the inhomogenous Teukolsky radial equation. We extend our analysis to arbitrary equatorial orbits in Section \ref{genericeqorbits} using conformal symmetry transformations. The main features of the waveforms for circular orbits are discussed in Section \ref{propobs} before concluding. 

%Tidal properties of neutron stars \cite{Damour:2009vw}t

%EFT approach  \cite{Bini:2015kja}\cite{Porto:2005ac,Porto:2008jj}\cite{Poisson:1997ha,Damour:2001tu}\cite{Tagoshi:1997jy} \cite{Steinhoff:2010zz,Steinhoff:2012rw,Steinhoff:2014kwa}\cite{Bohe:2015ana}\cite{Levi:2015msa}\cite{Goldberger:2007hy}. 

%For a review of the MPD formalism, see \cite{Dixon:2015vxa}

\section{The Mathisson-Papapetrou-Dixon formalism}

A point test particle with no internal structure and no gravitational self-force follows a geodesic path \cite{Einstein:1927aa}. The theory of motion of an extended object without gravitational self-force within a curved spacetime was initiated by Mathisson in 1937 \cite{Mathisson:1937zz}, rediscovered by Papapetrou in 1951 \cite{Papapetrou:1951pa} and completed beyond the dipole and quadrupole approximation by Dixon \cite{Dixon:1970zza,Dixon:1970zz,Dixon:1974aa}. It was also shown by Dixon \cite{Dixon:1964aa,Dixon:1970zza} and Schattner \cite{Schattner:1979vp,Schattner:1979vn} that the worldline of the center of mass of the extended object is uniquely defined from the Tulczyjew spin supplementary condition \cite{Tulczyjew:1959aa,Tulczyjew:1962aa}. A detailed physical and historical account can be found in \cite{Dixon:2015vxa}. We review here its most relevant features\footnote{Note that an alternative theory was proposed in \cite{Khriplovich:1989ed,Deriglazov:2017jub}. We shall not consider it here.}. 

\subsection{Evolution equations}

Neglecting its gravitational self-force, the motion of an extended body in a curved geometry is entirely determined by an infinite set of moment tensors defined along a timelike worldline within the object consisting of the momentum $p^\mu$, the antisymmetric spin tensor $S^{\alpha\beta}$, together with the quadrupole and higher order moment tensors. The Mathisson-Papapetrou-Dixon evolution equations of the momentum and spin are given by\footnote{We adopt the Misner-Thorne-Wheeler conventions for the Riemann tensor which are opposite to Dixon's. Our Riemann and Ricci tensor are $R^\rho_{\ \sigma\mu\nu}=\partial_{\mu} \Gamma_{\nu \sigma}^{\rho}-\partial_{\nu} \Gamma_{\mu \sigma}^{\rho}+\Gamma_{\mu \lambda}^{\rho} \Gamma_{\nu \sigma}^{\lambda}-\Gamma_{\nu \lambda}^{\rho} \Gamma_{\mu \sigma}^{\lambda}$ and $R_{\mu\nu} = R^\alpha_{\ \mu \alpha \nu}$.}
\bea
\frac{Dp^{\mu}}{D\tau}&=&-\frac{1}{2}R^{\mu}_{\;\,\nu\alpha\beta}u^{\nu}S^{\alpha\beta}+\mathcal{F}^{\mu},\label{eqn1}\\
\frac{DS^{\mu\nu}}{D\tau}&=&p^{\mu}u^{\nu}-p^{\nu}u^{\mu}+\mathcal{L}^{\mu\nu}\label{spin}
\eea
where $\mathcal{F}^{\mu}$ and  $\mathcal{L}^{\mu\nu}$ are respectively the force and the torque caused by the quadrupole and higher multipoles. The masses $\underbar m>0$, $m > 0$ are defined as
\bea
\underbar{m}^2 = -p^\mu p_\mu,\qquad m = -p^\mu u_\mu .
\eea

The $2^N$-pole moment, $N \geq 2$, is described by a tensor $J^{\mu_{1} \cdots \mu_{N-2} \alpha\beta\gamma\delta}$ with $N+2$ indices with the following symmetry structure
\bea
J^{\mu_{1} \cdots \mu_{N-2} \alpha\beta\gamma\delta} &=& J^{(\mu_{1} \mu_{N-2})[ \alpha\beta][\gamma\delta]} , \\
J^{\mu_{1}\cdots  \mu_{N-2} \alpha [\beta\gamma\delta]} &=& 0,\\
J^{\mu_{1}\cdots  \mu_{N-3}[\mu_{N-2} \alpha \beta]\gamma\delta} &=& 0, \qquad \text{for}\qquad N \geq 3.
\eea
Moreover, the octopole and higher order moments are defined with respect to an orthogonality relation involving a unit timelike vector $n^\alpha$ defined on the worldine,
\bea
n_{\mu_1} J^{\mu_{1} \cdots \mu_{N-2} \alpha\beta\gamma\delta} = 0, \qquad \text{for}\qquad N \geq 3. 
\eea
In particular the quadrupole $J^{\alpha\beta\gamma\delta}$ has $20$ independent components and has the symmetries of the Riemann tensor.

The force and torque can then be constructed as 
\bea
\mathcal{F}^{\mu}&=&\frac{1}{2}\sum_{N\ge 2}\frac{1}{N!}m^{\mu_1\cdots\mu_N\lambda\kappa}\nabla^{\mu}g_{\lambda\kappa,\mu_1\cdots\mu_N},\\
\mathcal{L}^{\mu\nu}&=&\sum_{N\ge 2}\frac{1}{(N-1)!}g^{\rho[\mu}m^{\nu]\mu_1\cdots\mu_{N-1}\alpha\beta}g_{\{\rho\mu,\nu\}\mu_1\cdots\mu_{N-1}}
\eea
where 
\be
m^{\mu_1\cdots\mu_N\rho\sigma\kappa}=\frac{4N}{N+2}J^{(\mu_1\cdots\mu_N|\sigma|\rho)\kappa},\qquad
g_{\{\alpha\beta,\gamma\}\delta\cdots}=g_{\alpha\beta,\gamma\delta\cdots}-g_{\beta\gamma,\alpha\delta\cdots}+g_{\gamma\alpha,\beta\delta\cdots}
\ee 
and $g_{\lambda\kappa,\mu_1\cdots\mu_N}$ is the $N$th extension of the metric $g_{\lambda\kappa}$ as defined by Veblen and Thomas \cite{Veblen:1923aa,Thomas:1934aa}. This is the unique tensor defined on each point $x^\mu_*(\tau)$ of the worldline whose components reduce to the $N$th partial derivative of the metric in any normal coordinate system centered on $x^\mu_*(\tau)$. The $N$th extension $g_{\lambda\kappa,\mu_1\cdots\mu_N}$ is symmetric in its indices $\mu_1 \dots \mu_N$.

\subsection{Choice of worldline as the center-of-mass}

The evolution equations need to be supplemented by a choice of worldline $x^\mu_*(\tau)$ within the extended body. As reviewed in \cite{Dixon:2015vxa}, one can uniquely identify the center-of-mass by imposing $n^{[\mu} p^{\nu]}=0$ (or more simply $n^\mu = p^\mu / (- p \cdot p)$) together with the Tulczyjew spin supplementary condition
\bea
S^{\mu\nu} p_\nu = 0. \label{SSC2}
\eea
We will assume the Tulczyjew spin supplementary condition from now on unless otherwise stated. Once these conditions are obeyed, the evolution equations \eqref{eqn1}-\eqref{spin} become the equations of motion of the momentum $p^\mu$ and spin tensor $S^{\alpha\beta}$, or equivalently, the spin vector 
\bea
S_\mu = \frac{1}{2 \, \underbar{m}}\eps_{\mu\nu\alpha\beta}p^\nu S^{\alpha\beta}. \label{spinv}
\eea
The spin length $S$ is then defined from either 
\bea
S^2 = \frac{1}{2}S_{\mu\nu}S^{\mu\nu} = S^\mu S_\mu.
\eea
We adopt the convention that the sign of $S$ is positive if the spin vector is aligned with the Kerr spin, and negative otherwise.
Note that the alternative spin supplementary condition originally assumed by Mathisson
\bea
S^{\mu\nu}u_\nu = 0\label{SSC3}
\eea 
does not uniquely fix a worldline already within special relativity \cite{Weyssenhoff:1947aa,Tulczyjew:1959aa}. Yet, it is often used in the literature instead of \eqref{SSC2}. %We will show in Section \ref{TODO} that assuming \eqref{SSC3} instead of \eqref{SSC2} would lead to a distinct gravitational wave spectrum. The validity of \eqref{SSC2} instead of \eqref{SSC3} could be therefore be assessed in principle by a gravitational wave observation. 

%\subsection{Solution for the velocity}
%
%The mass $\underbar{m}$ is defined as $\underbar{m}^2 = -p^\mu p_\mu$. Assuming \eqref{SSC2}, one can show that the velocity is then given by \cite{1977GReGr...8..197E}
%\bea
%u^\mu &=& h^\mu + \frac{1}{\underbar{m}^2-\frac{1}{4}R_{\alpha\beta\gamma\delta}S^{\alpha\beta}S^{\gamma\delta}}S^{\mu \nu}(\mathcal F_\nu -\frac{1}{2}R_{\nu \alpha\beta\gamma} h^\alpha S^{\beta\gamma})
%\eea
%where $h^\mu = \frac{p^\mu}{\underbar{m}} + \frac{1}{\underbar m^2}\mathcal L^{\mu\nu}p_\nu$. 

\subsection{Conserved quantities}

Given any Killing vector $\xi^\mu$ of the background spacetime, $\nabla_{(\mu} \xi_{\nu)} = 0$, one can show that the following quantity is conserved along the worldline, 
\bea
Q_\xi = \xi_\mu p^\mu + \frac{1}{2}S^{\mu\nu} \nabla_{\mu} \xi_{\nu} 
\eea
even in the presence of arbitrary multipoles  \cite{1977GReGr...8..197E}. This allows to define the probe energy and angular momentum uniquely.  Moreover, after chosing the quadrupole model \eqref{quadrupole}, the spin length $S$ is also conserved \cite{Steinhoff:2012rw}.

\subsection{Stress-tensor}

The Mathisson-Papapetrou-Dixon equations \eqref{eqn1}-\eqref{spin} can be shown to be equivalent to the conservation equations $\nabla_\nu T^{\mu\nu} = 0$ of the stress-energy tensor $T^{\mu\nu}$ described by the multipole moments. There is no further restriction on the multipole moments. 

In the dipole approximation (the test spinning particle), the stress-tensor is given by \cite{Mathisson2010,Papapetrou:1951pa,Tulczyjew:1957aa,Tulczyjew:1959aa,Blanchet:2013haa}
\be
T^{\mu\nu}=\int d\tau \{\frac{1}{2}(p^{\mu}u^{\nu}+p^{\nu}u^{\mu})\mathcal{D}-\frac{1}{2}\nabla_{\alpha}[(S^{\alpha\mu}u^{\nu}+S^{\alpha\nu}u^{\mu})\mathcal{D}]\}\label{stress}
\ee
where $\mathcal{D}$ is the Dirac function 
\be
\mathcal{D}=\frac{1}{\sqrt{-g}}\delta^{(4)}(x^{\mu}-x^{\mu}_*(\tau)).
\ee
In the quadrupole approximation, the stress tensor is given by \cite{Steinhoff:2009tk}
\be
T^{\mu\nu}=\int d\tau  [(p^{(\mu}u^{\nu)})\mathcal{D}+\frac{1}{3}R_{\alpha\beta\gamma}^{\hspace{14pt}(\mu}J^{\nu)\gamma\beta\alpha}\mathcal{D}-\nabla_{\alpha}(S^{\alpha(\mu}u^{\nu)}\mathcal{D})-\frac{2}{3}\nabla_{\alpha}\nabla_{\beta}(J^{\alpha(\mu\nu)\beta}\mathcal{D})].\label{Tmunu}
\ee

\subsection{Quadrupole model}

In the following we will consider exclusively the quadrupole approximation. The force and torque caused by the quadrupole $J^{\alpha\beta\gamma\delta}$ is more explicitly given by
\bea
\mathcal{F}^{\mu}&=&-\frac{1}{6}J^{\alpha\beta\gamma\delta}\nabla^{\mu}R_{\alpha\beta\gamma\delta},\\
\mathcal{L}^{\mu\nu}&=&\frac{4}{3} J^{\alpha\beta\gamma[\mu}R^{\nu]}_{\ \gamma\alpha\beta}.
\eea
We adopt a definition for the quadrupole $J^{\mu\nu\rho\sigma}$ which is a function of the momentum, spin tensor and Riemann curvature. The following three tensors share the symmetries of the Riemann tensor, 
\bea
Q_{(1)}^{\mu\nu\rho\sigma}&=&p^{[\mu}S^{\nu]}_{\ \alpha}S^{\alpha[\rho}p^{\sigma]},\\
Q_{(2)}^{\mu\nu\rho\sigma}&=&(p^{[\mu}R_{\alpha}^{\ \nu]\rho\sigma}+p^{[\rho}R_{\alpha}^{\ \sigma]\mu\nu})p^{\alpha},\\
Q_{(3)}^{\mu\nu\rho\sigma}&=&p^{[\mu}R^{\nu]\hspace{3pt}[\rho}_{\ \alpha\hspace{3pt}\beta}p^{\sigma]}p^{\alpha}p^{\beta}.
\eea
Following \cite{Steinhoff:2012rw}, we assume that the quadrupole tensor is a linear combination of these terms,
\bea
J^{\mu\nu\rho\sigma}&=&
\frac{m}{\underbar{m}^3}[\frac{3\kappa_{S^2}}{\underbar{m}}Q_{(1)}^{\mu\nu\rho\sigma}+\frac{3\mu_2+8\sigma_2}{\underbar{m}^2}Q_{(3)}^{\mu\nu\rho\sigma}-2\sigma_2Q_{(2)}^{\mu\nu\rho\sigma}]\label{J}\\
&=& \frac{m}{\underbar{m}^3}[\frac{3\kappa_{S^2}}{\underbar{m}}S^{\alpha[\mu}p^{\nu]}S_{\alpha}^{\  [\rho}p^{\sigma]}+3\mu_2 p^{[\mu}E^{\nu][\rho}p^{\sigma]}+2\sigma_2 (p^{[\mu}Q^{\nu]\rho\sigma}+p^{[\sigma}Q^{\rho]\nu\mu})].\nn
\label{quadrupole}
\eea
Here we defined 
\bea
Q^{\mu\nu\rho}&=&\epsilon^{\rho\nu}_{\ \ \alpha\beta}p^{\alpha}B^{\mu\beta},\\
E_{\mu\nu}&=&\frac{1}{\underbar{m}^2}R_{\mu\rho\nu\sigma}p^{\rho}p^{\sigma},\\
B_{\mu\nu}&=&\frac{1}{2\underbar{m}^2}\epsilon_{\mu\alpha\beta\gamma}R_{\nu\delta}^{\ \ \beta\gamma}p^{\alpha}p^{\delta}.
\eea
The coefficient $\kappa_{S^2}$ characterizes the spin-induced quadrupole. It is equal to $1$ for stationary black holes, which is equivalent to the quadrupole formula $Q=-J^2/M$ \cite{Geroch:1970cd,Hansen:1974zz}. For neutron stars, it depends upon its internal structure and is of the order of $\sim 5$ \cite{Laarakkers:1997hb}. The other two coefficients $\mu_2$ and $\sigma_2$ characterize the quadrupole deformations induced by the gravito-electric and gravito-magnetic tidal fields. They are approximately 0 for black holes and for neutron stars, they depend upon their internal structures.

We note the following dimensions of various quantities \footnote{We set the velocity of light $c=1$ and Newton constant $G_N = kg^{-2}$ such that space, time and inverse mass have the same dimension.  $``[O]"$ denotes the dimension of  $O$ in units of mass.},
\bea
&&[x^{\mu}]=-1,\ [\tau]=-1,\ [p^{\mu}]=1,\ [u^{\mu}]=0,\ [S^{\mu\nu}]=0,\ [g_{\mu\nu}]=0,\\&&[R^{\mu}_{\ \nu\rho\sigma}]=2,\ [J^{\mu\nu\rho\sigma}]=-1,
[G_N]=-2,\ [c]=0.
\eea
The dimension of the three parameters are then 
\be
[\kappa_{S^2}]=0,\ [\mu_2]=-3,\ [\sigma_2]=-3. 
\ee
The tidal deformation parameters are made dimensionless after introducing a typical radius scale $R$ (the horizon radius for a black hole). The dimensionless  tidal deformation parameters are usually defined as $k_2 = 3G\mu_2/(2R^5)$ and $j_2 = 48 G \sigma_2/R^5$. Realistic values for neutron stars are $k_2 \sim 0.1$, $j_2 \sim -0.02$ \cite{Flanagan:2007ix,Damour:2009vw,Binnington:2009bb}.

The masses $m$, $\underbar{m}$ are non-conserved and differ at $\mathcal O(S^3)$ in the spin tensor, 
\bea
m = \underbar{m} + O(S^3). 
\eea
We can therefore trade $\underbar{m}$ for $m$ in \eqref{J}. We also have $p^\mu = m u^\mu + \mathcal O(S^2)$.   The masslike quantity $\mu$ defined as 
\bea
\mu &=&m+\frac{\kappa_{S^2}}{2m}E_{\mu\nu}S^{\mu}_{\ \alpha}S^{\alpha\nu}+\frac{\mu_2}{4}E_{\mu\nu}E^{\mu\nu}+\frac{2}{3}\sigma_2B_{\mu\nu}B^{\mu\nu} \label{defmu} %\nn\\
%&=&m+\frac{\kappa_{S^2}}{2m^3}R_{\mu\rho\nu\sigma}S^{\mu}_{\ \alpha}S^{\alpha\nu}p^{\rho}p^{\sigma}+\frac{3\mu_2+8\sigma_2}{12m^4}R_{\mu\rho\nu\sigma}R^{\mu\ \ \nu}_{\ \rho'\ \sigma'}p^{\rho}p^{\sigma}p^{\rho'}p^{\sigma'}\nn\\
%&&+\frac{\sigma_2}{3m^2}R_{\alpha\mu\rho\sigma}R_{\beta}^{\ \mu\rho\sigma}p^{\alpha}p^{\beta}.
\eea
is conserved up to $\mathcal{O}(S^3)$. 

\subsection{Small mass ratio expansion}

We consider a small or intermediate mass ratio coalescence with 
\bea
q \equiv \frac{\mu}{M} \ll1. 
\eea
We assume that the probe object has a spin bounded by the extremal black hole bound. Equivalently, the intrinsic spin over mass square ratio
\bea
\chi \equiv \frac{S}{\mu^2}\quad \text{or, equivalently,}\quad S = \chi M^2 q^2\label{defchi}
\eea
obeys $-1 \leq \chi \leq 1$. In the $q$ expansion, the momentum is $p^\mu = O(q^1)$ and the spin tensor is $S^{\mu\nu} = O(q^2)$. The compactness coefficient of the probe $C \equiv \frac{G\mu}{R}$  as well as $k_2,j_2$ are assumed to be $O(q^0)$ in order to model black holes or neutron stars. We deduce that $\mu_2,\sigma_2 = O(q^5)$. The effects of electric or magnetic tidal deformations are therefore very negligeable for small $q$. We will not consider these effects furthermore in this paper. In the quadrupole model \eqref{J}, only the spin-induced quadrupole plays a role and it provides an $O(q^2)$ correction to the leading order result, as we will discuss.

Finally, we estimate the self-force (back-reaction) effects in terms of $q$. Self-force effects arise from the second (and higher) order curvature perturbations. Since the waveform is $O(q)$ at leading order, the self-force correction is $O(q^2)$, which is at the same order as the spin effects. Hence, our result below at linear order in the spin is only a subset of the relevant terms that describe the waveforms at order $O(q^2)$ and they would need to be completed with considerations of the self-force, which is beyond the scope of this paper.

\section{The Teukolsky problem for high spin IMRACs}
\label{HighSpinTeuk}

We set the background spacetime to be the nearly maximally spinning Kerr black hole of parameters $(M,J)$ in Boyer-Linquist coordinates $(\hat t,\hat r,\theta,\hat \phi)$ with 
\bea
\lambda = \sqrt{1-\frac{J^2}{M^4}} \ll 1. 
\eea
The high spin limit $\lambda \rightarrow 0$ leads to the existence of the NHEK region of coordinates $(T,R,\theta,\Phi)$ and the near-NHEK region of coordinates $(t,r,\theta,\phi)$ and parameter $\kappa$, both with enhanced $SL(2,\mathbb R) \times U(1)$ symmetry \cite{Bardeen:1999px,Amsel:2009ev,Bredberg:2009pv}. The NHEK metric is 
\be
ds^2=2M^2\Gamma(\theta) \left(-R^2 dT^2+\frac{dR^2}{R^2}+d\theta^2+\Lambda^2(\theta)(d\Phi+R dT)^2 \right),
\ee
where the polar functions are 
\be
\Gamma(\theta)=\frac{1+\cos^2\theta}{2},\qquad \Lambda(\theta)=\frac{2\sin\theta}{1+\cos^2\theta}.
\ee
The near-NHEK metric is 
\be
ds^2=2M^2\Gamma(\theta)\left( -r(r+2\kappa) dt^2+\frac{dr^2}{r(r+2\kappa)}+d\theta^2+\Lambda^2(\theta)(d\phi+(r+\kappa) dt)^2 \right).
\ee
%In the $\lambda \rightarrow 0$ limit,
 the change of coordinates between these patches is 
\bea
T=\frac{\hat t}{2M}\lambda^{2/3},\qquad R=\frac{\hat r - \hat r_+}{M}\lambda^{-2/3},\qquad \Phi = \hat \phi - \frac{\hat t}{2M},\label{chgt} \\
t =\frac{\hat t}{2M \kappa}\lambda,\qquad r=\kappa \frac{\hat r - \hat r_+}{M\lambda} ,\qquad \phi = \hat \phi - \frac{\hat t}{2M},\label{changecoord}
\eea
where $\kappa$ is arbitrary and factors out of any physical quantity.  Though $\kappa$ has no physical significance, we prefer to include it for bookkeeping purposes. In (near-)NHEK coordinates, the Kerr metric reduces to the (near-)NHEK metric plus $O(\lambda^{1/3})$ corrections. The ISCO lies within the NHEK region. For a review, see e.g. \cite{Compere:2018aar}. We will neglect such $O(\lambda^{1/3})$ corrections in this work: our results will be accurate at leading order in the high spin limit $\lambda \rightarrow 0$. We consider a source within the NHEK or the near-NHEK region. Circular orbits at a fixed Boyer-Linquist radius $\hat r = \hat r_0$ admit a fixed NHEK radius $R=R_0$ and near-NHEK radius $r= r_0$ determined from \eqref{chgt}-\eqref{changecoord} that is a functional of the constants of motion, see Section \ref{sec:nearorbit}.    Due to helicoidal symmetry, gravitational waves are sourced with a delta function $\delta(\hat \phi- \Omega_H \hat t)$ where $\Omega_H=1/(2M)+O(\lambda)$ is the angular velocity of the horizon. Within either of these regions, gravitational waves therefore have a frequency $\hat \omega \in \mathbb R$ and mode number $m \in \mathbb Z$ close to the superradiant bound $M \vert \hat \omega - \frac{m}{2M} \vert \ll 1$. We define the NHEK and near-NHEK angular velocities $\Omega$ and $\omega$ as 
\bea
\Omega = \frac{2M}{\lambda^{2/3}}(\hat \omega - \frac{m}{2M}), \qquad \omega =  \frac{2M\kappa}{\lambda}(\hat \omega - \frac{m}{2M}). 
\eea
At the leading order in the high spin limit, the Teukolsky perturbation can be deduced from a matched asymptotic expansion scheme. The main quantity of interest is the Newman-Penrose scalar $\delta \psi_4$ defined in the Kinnersley tetrad adapted to the Kerr geometry. We refer the reader to \cite{Compere:2017hsi} for a detailed derivation.  In the following two subsections, we will review how the observable large radius behavior of $\delta \psi_4$ can be expressed in terms of a series of coefficients $B_{lm}(x_*)$ which encode the details of the source. We will also present new formal identities (\ref{kappalimit}) and (\ref{ratioWN}) that allow to obtain the NHEK description in terms of a limit of the near-NHEK description.

\subsection{Source in the NHEK region}

For a source in the NHEK region, $\delta \psi_4$ asymptotes to
\bea
\delta \psi_4(\hat r \rightarrow \infty) =  \frac{M^3}{\sqrt{2 \pi}} \int_{-\infty}^{\infty} \d \Omega\, \sum_{lm} B_{lm} (x_*) \cK^{far}  S_{lm}(\theta)e^{im\hat{\phi}}e^{-\frac{i}{2M}(m+\lambda^{2/3}\Omega)\hat{u}} \hat{r}^{-1} \label{apsi4}
\eea
where the asymptotic retarded time $\hat{u} = \hat t - \hat r^*$ is defined in terms of the asymptotic tortoise coordinate $\hat r^*$, $S_{lm}(\theta)$ are the extremal spheroidal harmonics with separation constants $\mathcal E_{lm}=l(l+1)+O(m)$, and 
\bea
\cK^{far} &\equiv& \frac{(\lambda^{2/3})^{h} k_1}{1 -(-i\lambda^{2/3}\Omega)^{2h-1}k_2 } ,\label{cK}\\
k_1 &\equiv&  \frac{2^{im}e^{-im/2}\Gamma(2-2h)}{\Gamma(1-h+im-s)} (im)^{h-1+im-s} \left[1-\frac{(-im)^{2h-1}}{(im)^{2h-1}} \frac{\sin{\pi(h+im)}}{\sin{\pi(h-im)}} \right], \\
k_2 &\equiv& (-2i  m)^{2h-1} \frac{\Gamma(1-2h)^2}{\Gamma(2h-1)^2} \frac{\Gamma(h-im+s)}{\Gamma(1-h-im+s)} \frac{\Gamma(h-im-s) }{\Gamma(1-h-im-s)},\\
h &\equiv& \frac{1}{2} (1+ \text{sign}(\eta^2_{lm}) \eta_{lm}), \qquad \eta_{lm}\equiv\sqrt{1-7m^2 + 4 \mathcal E_{lm}}.
\eea
Here $s=-2$ is understood, but the expression generalizes to other spins. 

The coefficients $B_{lm}(x_*)$ encode the details of the source. In the NHEK region, the homogeneous solution to the radial equation outside of the source
\bea
R_{lm\Omega}(R) |_{R>R_*(\tau)} &=&  A_{lm}(x_*)\, \cW^{\text{in}}_{lm\Omega}(R) +B_{lm}(x_*) \,  \cM^{\text{D}}_{lm\Omega}(R) \label{Rexo}
\eea
is a linear combination of two Whittaker functions\footnote{With respect to \cite{Compere:2017hsi}, we changed the normalization by a factor of $ (-2 i \Omega)^{-h}$. All other conventions are unchanged.}
\begin{eqnarray}
\cW^{\text{in}}_{lm\Omega}(R) &\equiv& (-2 i \Omega)^{-h} R^{-s}W_{im+s,h-1/2}(\frac{-2i \Omega}{R}) ,\nn\\
\cM^{\text{D}}_{lm\Omega}(R) & \equiv& (-2 i \Omega)^{-h}  R^{-s} M_{im+s,h-1/2}(\frac{-2i \Omega}{R}). \label{Rnn}
\end{eqnarray}
We write the solution inside of the source as
\bea
R_{lm\Omega}(R) |_{R<R_*(\tau)} &=&  (A_{lm}(x_*) +C_{lm}(x_*))\, \cW^{\text{in}}_{lm\Omega}(R).
\eea
The Wronskian is 
\bea
\mathcal W \equiv- (- 2i\Omega)^{1-2h}\frac{\Gamma(2h)}{\Gamma(h-im-s)}.
\eea
Outgoing boundary conditions imply
\bea
A_{lm}=B_{lm} \frac{\Gamma(1-h-i m -s)}{\Gamma(1-2h) (\frac{(-i\lambda^{2/3}\Omega)^{1-2h}}{k_2} -1)}.\label{Alm}
\eea
The problem of solving the Teukolsky problem at the leading order in the limit $\lambda \rightarrow 0$ is therefore reduced to the problem of finding the $B_{lm}(x_*)$ coefficients in the NHEK solution \eqref{Rexo}. 

\subsection{Source in the near-NHEK region}

For a source in the near-NHEK region, $\delta \psi_4$ asymptotes to
\be
\delta \psi_4(\hat{r} \to \infty) =  \frac{M^3}{\sqrt{2 \pi}} \int_{-\infty}^{\infty} \d \omega \,  \sum_{lm} B_{lm}(x_*)  \cK^{far}_{\kappa}  S_{lm}(\theta)e^{im\hat{\phi}}e^{-i\hat{\omega} \hat u}  \hat{r}^{-1}
\label{eqn:nearpsi4}
\ee
with
\bea
\cK^{far}_{\kappa} &\equiv  &  \frac{  \lambda^{h} \kappa^{-h} k_1}{ 1 - \lambda^{2h-1} k_2 \frac{\Gamma(h-in+im)}{\Gamma(1-h-in+im)}},\label{Kk} \\ 
n &\equiv & \frac{\omega}{\kappa} +m.\label{defn}
\eea
The source-dependent coefficient $B_{lm}(x_*)$ is determined from the homogenous solution to the radial equation in near-NHEK
\be
	R_{lm\omega}(r)|_{r>r_{*}(\tau)} =  A_{lm}(x_*) \cR_{lm\omega}^{\text{in}}(r) + B_{lm}(x_*) \cR_{lm\omega}^{\text{D}}(r)  
	\label{eqn:nearNHEKmatching}
\ee
where the two independent solutions are the following hypergeometric functions
\bea
	\cR_{lm\omega}^{\text{in}}(r) &=& r^{-in/2-s}(\frac{r}{2\kappa}+1)^{i(\frac{n}{2}-m)-s} {}_2F_1(h-im-s,1-h-im-s,1-in-s,-\frac{r}{2 \kappa}) ,\nn\\
	\cR_{lm\omega}^{\text{D}}(r) &=& r^{-h-s}(\frac{2\kappa}{r}+1)^{i(\frac{n}{2}-m)-s} {}_2F_1(h-im-s,h-im+in,2h,-\frac{2 \kappa}{r}). \label{Rnnn}
\eea
The Wronskian is 
\bea
\mathcal W_{\kappa} \equiv -\frac{(2\kappa)^{1-h-in/2}\Gamma(2h)\Gamma(1-in-s)}{\Gamma(h+im-in)\Gamma(h-im-s)}.\label{WnNHEK}
\eea
The problem is again reduced to finding the $B_{lm}(x_*)$ coefficients in the near-NHEK solution \eqref{eqn:nearNHEKmatching}. 

In order to relate the expressions in near-NHEK to the NHEK expressions, note the following limit,  
\bea
\lim_{\kappa \rightarrow 0} \kappa^{2h-1}\frac{\Gamma(h-i n + i m)}{\Gamma(1-h-i n+im)}=(-i\omega)^{2h-1}. \label{kappalimit}
\eea
It implies the following formal limit
\bea
\mathcal K^{far} = \lim_{\kappa = \lambda^{1/3}\rightarrow 0} \mathcal K^{far}_\kappa \Big|_{\omega \mapsto \Omega}. 
\eea
It is then straightforward to show that the near-NHEK and NHEK solutions are related as 
\bea
\frac{\mathcal W^{\text{in}}_{lm\Omega}}{\mathcal W } = \lim_{\kappa \rightarrow 0} \frac{\mathcal R_{lm\omega}^{\text{in}}}{\mathcal W_{\kappa}} \Big|_{\omega \mapsto \Omega}. \label{ratioWN}
\eea
We can therefore only obtain the near-NHEK expressions, and perform the above limit to obtain the NHEK results.

\section{The circular (near-)NHEK orbit}
\label{sec:nearorbit}

We consider a circular orbit in near-NHEK spacetime. The orbit is characterized by its near-NHEK energy $e$ and angular momentum $\ell$ per unit probe mass $\mu = M q$. For a circular orbit in NHEK spacetime, one can set $\kappa = 0$ and switch notation as $r \mapsto R$, $t \mapsto T$, $\phi \mapsto \Phi$, $e \mapsto E$ in order to obtain the explicit expressions in NHEK spacetime. We will explicate most formulae only for the near-NHEK orbit. 

The near-NHEK trajectory is
\bea
r= r_0,\qquad \theta = \frac{\pi}{2},\qquad \phi = - \alpha r_0 t
\eea
where $\alpha$ is the rescaled angular velocity. The normalization condition $u^2=-1$ solved for 
\bea
u^t = \frac{1}{M r_0 \sqrt{8(1+\kappa_0)\alpha - (3+4\alpha^2 +6\kappa_0+4\kappa_0^2)}},\qquad u^\phi = -\alpha r_0 u^t,
\eea
where we defined $\kappa_0 \equiv \frac{\kappa}{r_0}$. 

The NHEK trajectory is
\bea
R=R_0,\qquad \theta = \frac{\pi}{2},\qquad \Phi = - \alpha R_0 T
\eea
with 
\bea
u^T = \frac{1}{M R_0 \sqrt{-3+8\alpha - 4\alpha^2}},\qquad u^\Phi = -\alpha R_0 u^T. 
\eea

\subsection{Solution to the MPD equations}

 We choose the orientation $\eps_{tr \theta\phi}=+1$. Assuming \eqref{SSC2} and using the definitions of the probe mass \eqref{defmu} and spin ratio \eqref{defchi}, we obtain the following near-NHEK solution to the MPD equations
\bea
S^{tr}&=&\frac{(1+\kappa_0)\chi q^2}{\lambda_0}(1+\frac{6(1+2\kappa_0)\chi q}{\lambda_0^2}+\mathcal{O}(q^2)),\\
S^{r\phi}&=&\frac{r_0\kappa_0^2\chi q^2}{\lambda_0}(1+\frac{9(1+\kappa_0)^2(1+2\kappa_0)\chi q}{2\kappa_0^2\lambda_0^2}+\mathcal{O}(q^2)),\\
p^t&=&\frac{2q}{r_0\lambda_0}(1+\frac{3(1+\kappa_0)^2\chi q}{2\lambda_0^2}\nn\\&&+\frac{(3(1+\kappa_0)^2(6+12\kappa_0+\kappa_0^2)+2(-9+\kappa_0(-36-36\kappa_0+\kappa_0^3))\kappa_{S^2})\chi^2q^2}{2\lambda_0^4}+\mathcal{O}(q^3)),\nn
\eea
\bea
p^{\phi}&=&-\frac{3(1+\kappa_0)q}{2\lambda_0}(1+\frac{2\kappa_0^2\chi q}{\lambda_0^2}\nn\\&&+\frac{(2\kappa_{S^2}(-9-36\kappa_0-36\kappa_0^2+\kappa_0^4)+9+36\kappa_0+57\kappa_0^2+42\kappa_0^3+4\kappa_0^4)\chi^2q^2}{2\lambda_0^4}+\mathcal{O}(q^3)),\nn\\
\alpha&=&\frac{3}{4}(1+\kappa_0)(1-\frac{\chi q}{2}+\frac{1}{4}(4\kappa_{S^2}-5)\chi^2q^2+\mathcal{O}(q^3))\label{defalphaNEARNHEK}
\eea
where $\lambda_0=\sqrt{3+6\kappa_0-\kappa_0^2}$. The specific energy, angular momentum and non-invariant mass are given by 
\bea
e&\equiv& \frac{Q_{-\p_t}}{\mu} = -\frac{2Mr_0\kappa_0^2}{\lambda_0}(1+\frac{(9+18\kappa_0+\kappa_0^2)\chi q}{2\lambda_0^2}\nn\\&&+\frac{(27(1+\kappa_0)^2(1+2\kappa_0)+2(-9+\kappa_0(-36-36\kappa_0+\kappa_0^3))\kappa_{S^2})\chi^2q^2}{2\lambda_0^4}+\mathcal{O}(q^3)),\\
\ell&\equiv& \frac{Q_{\p_\phi}}{\mu} =\frac{2M(1+\kappa_0)}{\lambda_0}(1+\frac{(3+\kappa_0(6+\kappa_0))\chi q}{\lambda_0^2}\nn\\&&+\frac{(2\kappa_{S^2}(-9-36
\kappa_0-36\kappa_0^2+\kappa_0^4)+9+36\kappa_0+69\kappa_0^2+66\kappa_0^3)\chi^2q^2}{2\lambda_0^4}+\mathcal{O}(q^3)),\\
\underbar{m}&=&Mq(1-\frac{(3+6\kappa_0+\kappa_0^2)\kappa_{S^2}\chi^2q^2}{\lambda_0^2}+\mathcal{O}(q^3)).
\eea
  The only non-vanishing component of the spin vector is $S_\theta = -M^3 \chi q^2 + O(q^4)$. Since the direction of $S_z$ and $S_{\theta}$ are opposite for an equatorial orbit, the probe spin vector is aligned with the Kerr spin for $\chi >0$ and anti-aligned for $\chi < 0$. We checked that the solution is the unique perturbation of the spinless equatorial circular orbit. 
This solution matches the leading order high spin limit $\lambda \rightarrow 0$ of the solution provided in Section IV.A. of \cite{Hinderer:2013uwa} after specializing to probe black holes which have $\kappa_{S^2}=1$ and after switching the sign convention for the definition of $S^\mu \rightarrow S^\mu_{their}=- S^\mu$ in \eqref{spinv}. More precisely, our $q \chi$ matches with their $-\text{sign}(S_* L_z)S_*$. Therefore, their probe spin vector is aligned with the Kerr spin for $\chi <0$ and anti-aligned for $\chi > 0$. The orbit considered is prograde relative to the Kerr spin, $\text{sign}(a L_z) =+1$. Note that all quantities in \cite{Hinderer:2013uwa} are given in Boyer-Linquist coordinates while ours are given in near-NHEK coordinates. They are related by \eqref{chgt}-\eqref{changecoord}. 

We note the relation 
\bea
e=-\frac{\sqrt{3}\kappa}{2}\sqrt{\ell^2-\ell_*^2} \left( 1+\frac{1}{2}(\kappa_{S^2}-1)\chi^2 q^2 \right)+O(q^3).\label{defe}
\eea
as well as 
\bea
p^t&=&\underbar{m} u^t+\frac{6(1+\kappa_0)^2(\kappa_{S^2}-1)\chi^2 q^3}{r_0\lambda_0^3}+\mathcal{O}( q^4),\label{pt}\\
p^{\phi}&=&\underbar{m} u^{\phi}-\frac{6\kappa_0^2(1+\kappa_0)(\kappa_{S^2}-1)\chi^2 q^3}{\lambda_0^3}+\mathcal{O}( q^4).\label{pphi}
\eea
Therefore, the momentum is not aligned with $\underbar{m} u^\mu$ due to $ q^3$ corrections which are also quadratic in the spin. In the special case where the compact object is a black hole, $\kappa_{S^2}=1$, $p^\mu = \underbar{m} u^\mu + O( q^4)$ and the spin supplementary conditions $p^\mu = \underbar{m} u^\mu$ and \eqref{SSC3} are equivalent to \eqref{SSC2} at this order. Another solution to the MPD equations for circular orbits appeared in \cite{Bini:2015zya}. Their model is exactly the same as ours, including the supplementary condition. Therefore our solution is the large spin limit of the one  in \cite{Bini:2015zya}. Indeed,  we checked that the velocity, momentum, spin, energy and angular momentum of the equatorial circular orbits match with our solution for general $\kappa_{S^2}$ up to $\mathcal{O}(q^2)$ in the high spin limit. From \cite{Bini:2015zya}, we also infer that the ISCO radius is given by 
\be
R_0=2^{1/3}+2^{4/3}\chi q+\frac{27-16\kappa_{S^2}}{2^{5/3}}(\chi q)^2+O(q^3)\label{ISCOspin}
\ee
in the high spin limit.

In NHEK spacetime, the specific energy and angular momentum are given by 
\bea
E=0,\qquad \ell = \ell_*[\chi q] \equiv \frac{2M}{\sqrt{3}}(1+\chi q + (\frac{1}{2}-\kappa_{S^2})(\chi q)^2)+O(q^3) \label{lstar}
\eea
while the radius $R_0$ is fixed as \eqref{ISCOspin}. This defines the critical specific angular momentum $\ell_*$ as function of the reduced intrinsic spin $\chi q = \frac{S}{\mu M}$. This generalizes the point particle definition $\ell_*[0] = \frac{2}{\sqrt{3}} M$  \cite{Compere:2017hsi}. Following the change of coordinates between the NHEK coordinates and Boyer-Linquist coordinates, the critical angular momentum can be defined in terms of the asymptotically flat energy $\hat E_{ISCO}$ of the ISCO as 
\bea
\ell_* = \frac{\hat E_{ISCO}}{\Omega_{ext}}
\eea
where the black hole angular velocity at extremality is $\Omega_{ext}=(2 M)^{-1}$.

\subsection{Reparametrization in terms of invariants}

For a probe particle we can invert the relation between $e,\ell$ and $\kappa_0,r_0$ to obtain 
\bea
\kappa_0 = \kappa_{00}[\ell] \equiv   \left( \frac{2\ell}{\sqrt{3(\ell^2 - \ell^2_{*}[0])} }-1 \right)^{-1},\qquad r_0= r_{00}[e,\ell] \equiv -\frac{e}{\ell} \frac{1+\kappa_{00}[\ell]}{\kappa_{00}^2[\ell]}.
\eea
After including the multipole moments, we instead have
\bea
\kappa_0=\kappa_{00} (1+\sum_{i\ge1}\kappa_{0i}q^i), \qquad  r_0=r_{00}(1+\sum_{i\ge1}r_{0i}q^i) \label{r0}
\label{kappa0x}
\eea
which we can solve iteratively for the correction coefficients. We find the convenient form 
\bea
\kappa_0 &=&  \left(-1+ \frac{2\ell}{\sqrt{3(\ell^2 - \ell^2_*)} } (1+\frac{\chi q}{2}+(1-\frac{\kappa_{S^2}}{2})\chi^2 q^2 ) \right)^{-1}+O(q^3),\label{k0d} \\
r_0 &=& -\frac{e}{\ell} \frac{1+\kappa_0}{\kappa_0^2} ( 1-\frac{\chi q}{2}-\frac{\chi^2 q^2}{4})+O(q^3). \label{r0f}
\eea
%\bea
%r_{01}&=&\frac{6+15\kappa_{00}+6\kappa_{00}^2+\kappa_{00}^3}{4\kappa_{00}^2}\chi,\nn\\
%\kappa_{01}&=&-\frac{(1+\kappa_{00})(3+6\kappa_{00}+\kappa_{00}^2)}{4\kappa_{00}^2}\chi,\nn\\
%r_{02}&=&\frac{\chi^2}{32\kappa_{00}^4}(72 + 333 \kappa_{00} + 492 \kappa_{00}^2 + 246 \kappa_{00}^3 + 68 \kappa_{00}^4 + 21 \kappa_{00}^5 \\
%&& -8\kappa_{S^2} \kappa_{00}^2 (6 + 15 \kappa_{00} + 8 \kappa_{00}^2 + \kappa_{00}^3)),\nn\\
%\kappa_{02}&=&\frac{\chi^2(1+\kappa_{00})}{32\kappa_{00}^4}(-9 - 18 \kappa_{00} + 30 \kappa_{00}^2 + 72 \kappa_{00}^3 + 3 \kappa_{00}^4 + 2 \kappa_{00}^5+8\kappa_{S^2}\kappa_{00}^2(3+6\kappa_{00}+\kappa_{00}^2)).\nn
%\eea 
After some algebra, we can rewrite the solution in a simple form in terms of invariants as 
\bea
S^{tr}&=&\frac{\ell\chi q ^2}{\sqrt{3}\ell_*}(1+2 \chi q )+O(q^4),\\
S^{r\phi}&=&-\frac{e\chi q ^2}{\sqrt{3}\ell_*}( 1+ \frac{2\chi q}{1-\frac{\ell_*^2}{\ell^2}})+O(q^4),\label{pt1}\\
p^t&=&-\frac{\sqrt{3} \ell_*q}{2e} (\frac{\ell^2}{\ell_*^2}-1)(1-\frac{\chi^2 q^2}{2})+O(q^4),\label{pphi1}\\
p^{\phi}&=&-\frac{\sqrt{3} \ell q }{2\ell_*} (1 +(\frac{1}{2}-\kappa_{S^2}) \chi^2q ^2)+O(q^4),\\
%\alpha&=&\frac{3}{4}(1+\kappa_{00})(1-\frac{\chi q }{4\kappa_{00}}(3+8\kappa_{00}+\kappa_{00}^2)+\frac{\chi^2q ^2}{32\kappa_{00}^3}(8\kappa_{S^2}\kappa_{00}^2 (3 + 10 \kappa_{00} + \kappa_{00}^2)\nn\\&& +(1+\kappa_{00})(-1+2\kappa_{00})(9+27\kappa_{00}+3 \kappa_{00}^2+\kappa_{00}^3) )+O(q^3),\label{alphax}\\
\underbar m&=&Mq (1-\frac{\kappa_{S^2}}{2} (\frac{\ell^2}{\ell_*^2}+1)\chi^2 q^2)+O(q^4),\\
\frac{\alpha}{\kappa_0} &=& \frac{\sqrt{3} \ell}{2 \sqrt{\ell^2 - \ell_*^2}}(1 + \frac{1}{2}(\kappa_{S^2}-1)\chi^2 q^2)+O(q^3). \label{alphakappa}
\eea
In the NHEK case, the analogue solution reads
\bea
S^{tr}&=&\frac{\chi q ^2}{\sqrt{3}}(1+2 \chi q )+O(q^4),\\
S^{r\phi}&=&\frac{\sqrt{3}R_0 \chi^2 q^3}{2}+O(q^4),\label{pt2}\\
p^t&=&\frac{2q}{\sqrt{3}R_0}(1+\frac{1}{2}\chi q + (1-\kappa_{S^2})\chi^2 q^2)+O(q^4),\label{pphi2}\\
p^{\phi}&=&-\frac{\sqrt{3} q }{2} (1 +(\frac{1}{2}-\kappa_{S^2}) \chi^2q ^2)+O(q^4),\\
\underbar m&=&Mq (1-\kappa_{S^2}\chi^2 q^2)+O(q^4),\\
\alpha&=&\frac{3}{4} (1-\frac{\chi q}{2}+(\kappa_{S^2}-\frac{5}{4})\chi^2q^2+\mathcal{O}(q^3)).\label{defalpha}
\eea
Formally, one can obtain the NHEK result from the near-NHEK result by defining the NHEK radius $R_0$ as 
\bea
\kappa = \frac{\sqrt{3}R_0 }{2} \sqrt{\frac{\ell^2}{\ell^2_*} - 1}\label{defR0}
\eea
and performing the limit $\ell \rightarrow \ell_*$ at $R_0$ fixed. The near-NHEK energy $e$ defined in \eqref{defe} then vanishes in the limit, which gives the NHEK energy $E=0$. The substitution of the near-NHEK energy combined with the definition \eqref{defR0} allows to cancel the pole and zero in \eqref{pt1}-\eqref{pphi1} and to correctly recover \eqref{pt2}-\eqref{pphi2}. 

\subsection{Stress-tensor}

It is clear from the explicit solution of the circular near-NHEK orbit that $p^\mu$, $S^{\mu\nu}$ are independent of the coordinates. We can therefore straightforwardly integrate \eqref{Tmunu} using $t$ as time on the worldline. There are maximally $N$ derivatives acting on the delta functions when the $2^N$ multipoles are included. This leads to an expression of the form 
\bea
T^{\mu\nu} = \sum_{i,j,k \geq 0 }^{i+j+k \leq N} T^{\mu\nu}_{ijk}\delta^{(i)}(r-r_0) \delta^{(j)}(\theta - \frac{\pi}{2})\delta^{(k)} (\phi + \alpha r_0 t). 
\eea
Following the notations of \cite{Compere:2017hsi}, the right-hand side of the Teukolsky equation is given for spin $s=2$ by $T_{(2)}=-2\mathcal T_0$ and for spin $-2$ by $T_{(-2)} = -2 \rho^{-4}\mathcal T_4$ where $\rho=-(\hat r - i a \cos\theta)^{-1}$ and $\mathcal T_0$ and $\mathcal T_4$ are given in (A.29)-(A.30) of \cite{Compere:2017hsi}. This leads to
\bea
\mathcal T_4 &=& \sum_{i,j,k \geq 0 }^{i+j+k \leq N+2} t^4_{ijk}\delta^{(i)}(r-r_0) \delta^{(j)}(\theta - \frac{\pi}{2})\delta^{(k)} (\phi + \alpha r_0 t),\label{T4}\\
\mathcal T_0 &=& \sum_{i,j,k \geq 0 }^{i+j+k \leq N+2} t^0_{ijk}\delta^{(i)}(r-r_0) \delta^{(j)}(\theta - \frac{\pi}{2})\delta^{(k)} (\phi + \alpha r_0 t).  \label{T0}
\eea

The Teukolsky equation with source terms (\ref{T4}) and (\ref{T0}) can be solved by separation of variables. The angular part of the waveform satisfies the differential equation 
\be
\frac{1}{\sin \theta} \frac{\mathrm{d}}{\mathrm{d} \theta}\left(\sin \theta \frac{\mathrm{d} S_{l m}}{\mathrm{d} \theta}\right)+\left[\frac{m^{2}}{4} \cos ^{2} \theta-m s \cos \theta-\left(\frac{m^{2}+2 m s \cos \theta+s^{2}}{\sin ^{2} \theta}\right)+\mathcal{E}_{l m}\right] S_{l m}=0 \label{angular}
\ee 
 in which the separation constant $\mathcal{E}_{l m}$ is independent of $\omega$. The disappearence of the frequency in this angular equation is due to the locking of the frequency in terms of the angular mode $m$ due to the kinematics of the near-horizon region: $\hat \omega = \frac{m}{2M}+O(\lambda)$. The angular equation can be obtained either from the Teukolsky equation in Kerr at leading order in the high spin limit or directly from the Teukolsky equation in the NHEK or near-NHEK background, see Appendix A of \cite{Compere:2017hsi} for details. Since the separation constant is independent from $\omega$, it can be solved numerically independently from the radial problem. Its solutions are called the extremal spin-weighted spheroidal harmonic functions.

To solve the radial Teukolsky equation, we need to expand the stress-tensor source in terms of  the extremal spin-weighted spheroidal harmonic functions as
\bea
\mbox{}_{-2}T_{lm\tilde{\Omega}}(r)&=&-4M^2\int_{0}^{2\pi} d\phi e^{-im(\phi-\tilde{\omega}t)}\int_0^{\pi}d\theta \sin\theta S_{lm}(\theta)(1+\cos^2\theta)(1-i\cos\theta)^4 \mathcal T_4\nn\\
&=&q\sum_{i=0}^{N+2} \mbox{}_{-2}t_i\,  \frac{r_0^{i+3}}{M^4} \delta^{(i)}(r-r_0); \label{Tm2d}\\
\mbox{}_{+2}T_{lm\tilde{\Omega}}(r)&=&-4M^2\int_{0}^{2\pi} d\phi e^{-im(\phi-\tilde{\omega}t)}\int_0^{\pi}d\theta \sin\theta S_{lm}(\theta)(1+\cos^2\theta) \mathcal T_0\nn\\
&=&q\sum_{i=0}^{N+2} \mbox{}_{+2}t_i \, r_0^{i-1}\delta^{(i)}(r-r_0). \label{Tp2d}
\eea
The computation of the stress-tensor including up to quadrupole moment contributions amounts to finding the 10 coefficients $\mbox{}_{s}t_i$ for $s=\pm 2$, $i=0,\dots 4$. Note that upon substituting $\kappa = r_0 \kappa_0$ and expressing all quantities in terms of $r_0,\kappa_0,M$, the powers of $r_0$ and $M$ factor in the stress-tensor. We defined the coefficients $\mbox{}_{s}t_i$ so that they only depend upon the product $q \chi$, $\kappa_0$ and $m$. Explicitly, we need to compute 
\bea
q \mbox{}_{-2}t_{\text{i}}\,  r_0^{\text{i}+3} \hspace{-9pt}&=& \hspace{-9pt}-4M^6 \sum_{k \geq 0}^{N+2-\text{i}} (i m)^k \sum_{j \geq 0}^{N+2-\text{i}-k} (-1)^j \frac{\p^j}{\p \theta^j}[\sin\theta S_{lm}(\theta)(1+\cos^2\theta)(1-i\cos\theta)^4 t^4_{\text{i} jk}]\vert_{\theta=\frac{\pi}{2}} ;\nn \\
q \mbox{}_{+2}t_\text{i}\,  r_0^{\text{i}-1}  \hspace{-9pt}&=&  \hspace{-9pt}-4M^2 \sum_{k \geq 0}^{N+2-\text{i}} (i m)^k \sum_{j \geq 0}^{N+2-\text{i}-k} (-1)^j \frac{\p^j}{\p \theta^j}[\sin\theta S_{lm}(\theta)(1+\cos^2\theta) t^0_{\text{i} jk}]\vert_{\theta=\frac{\pi}{2}} .
\eea
The result is displayed in Appendix \ref{coefst4t0}. The coefficients for the NHEK case are simply obtained by setting $\kappa_0 = 0$.

\subsection{Solution to the radial Teukolsky equation}

For circular orbits in near-NHEK equatorial plane, the radial Teukolsky differential equation for $\psi_4$ takes the form \eqref{deom} with \eqref{ABn} where the source $T(r)= \mbox{}_{-2}T_{lm\tilde{\Omega}}(r)$ is a sum of derivatives of delta functions as obtained in \eqref{Tm2d}. Our goal is to find the $B_{lm}$ coefficients in \eqref{eqn:nearNHEKmatching} as a function of the source parameters, namely the specific angular momentum $\ell$, the specific near-NHEK energy $e$, the mass ratio $q$ and the intrinsic spin over mass square ratio $\chi$. The black hole mass $M$ gives the overall scale. The spin $J$ does not appear in $B_{lm}$ since we already truncated the Teukolsky perturbation by keeping only the leading order contribution close to maximal spin for each $l,m$ in \eqref{eqn:nearpsi4}.

The $B_{lm}$ coefficients are exactly the $X_2$ coefficients of the general solution \eqref{dif} given in \eqref{X2coef} with $a_i = q M^{-4}r_0^{i+3}\mbox{}_{-2}t_{{i}}$. 
Explicitly, we find 
\bea
\hspace{-17pt}B_{lm}(x_*)\hspace{-7pt}&=\hspace{-7pt}& \frac{q}{\mathcal W_{\kappa} M^4 r_0} \Big\{ \frac{\mathcal R_{lm\omega}^{\text{in}}(r_0)}{(1+2\kappa_0)^2} \Big[ \mbox{}_{-2}t_{4} \Big(\frac{V^2(r_0)}{(1+2\kappa_0)^2} +\frac{r_0^2 V''(r_0)}{1+2\kappa_0} \Big) - b_3 \frac{r_0 V'(r_0)}{1+2\kappa_0}  \nn\\
&& +b_0+b_2 \frac{V(r_0)}{1+2\kappa_0}  \Big] +\frac{r_0 \mathcal R_{lm\omega}^{\text{in}\prime}(r_0) }{(1+2\kappa_0)^2} \Big[ -b_1 +2 \mbox{}_{-2}t_{4} \frac{r_0 V'(r_0)}{1+2\kappa_0}-\tilde b_3 \frac{V(r_0)}{1+2\kappa_0} \Big] \Big\}\label{Blm}
\eea
where
\bea
b_0 &=&\mbox{}_{-2}t_{0}+ 4 \frac{1+\kappa_0}{1+2\kappa_0} \mbox{}_{-2}t_{1}+4  \frac{5+10\kappa_0+6 \kappa_0^2}{(1+2\kappa_0)^2} \mbox{}_{-2}t_{2}+ 24\frac{(1+\kappa_0)(5+10\kappa_0+8\kappa_0^2)}{(1+2\kappa_0)^3} \mbox{}_{-2}t_{3} \nn\\
&&+24\frac{35+140 \kappa_0+252 \kappa_0^2+224\kappa_0^3+80\kappa_0^4}{(1+2\kappa_0)^4} \mbox{}_{-2}t_{4}  ,\nn\\
b_1 &=&\mbox{}_{-2}t_{1}+ 6 \frac{1+\kappa_0}{1+2\kappa_0} \mbox{}_{-2}t_{2}+2  \frac{19+38\kappa_0+24\kappa_0^2}{(1+2\kappa_0)^2} \mbox{}_{-2}t_{3}+ 16\frac{(1+\kappa_0)(17+34\kappa_0+30\kappa_0^2)}{(1+2\kappa_0)^3} \mbox{}_{-2}t_{4} ,\nn\\
b_2 &=& \mbox{}_{-2}t_{2}+12 \frac{1+\kappa_0}{1+2\kappa_0} \mbox{}_{-2}t_{3}+2\frac{61+122\kappa_0+72 \kappa_0^2}{(1+2\kappa_0)^2} \mbox{}_{-2}t_{4},\\
b_3 &=& \mbox{}_{-2}t_{3} + 18 \frac{1+\kappa_0}{1+2\kappa_0} \mbox{}_{-2}t_{4},     \nn   \\
\tilde b_3 &=& \mbox{}_{-2}t_{3} + 16 \frac{1+\kappa_0}{1+2\kappa_0} \mbox{}_{-2}t_{4}.  \nn   
\eea
The coefficients $\mbox{}_{-2}t_{i}$ are written in Appendix \ref{coefst4t0} and the Wronskian $\mathcal W_\kappa$ is given in \eqref{WnNHEK}. 

The final Teukolsky perturbation takes the form 
\be
\delta \psi_4(\hat{r} \to \infty) =  \frac{1}{\hat r}  \sum_{l,m} M^3 B_{lm}(x_*)  \cK^{far}_{\kappa}  S_{lm}(\theta)e^{im\hat{\phi}-i\hat{\omega} \hat u}  
\ee
where $\cK^{far}_{\kappa}$ is defined in \eqref{defn} and $n=m(1-\frac{\alpha}{\kappa_0})$. We define $\tilde{\mathcal R}_{lm\omega}^{\text{in}}(\kappa_0)$ from
\bea
\mathcal R_{lm\omega}^{\text{in}}(r_0) &=& r_0^{-\frac{in}{2}-s} \tilde{\mathcal R}_{lm\omega}^{\text{in}}(\kappa_0). 
%\mathcal R^{D}(r_0) = r_0^{- h-s} \tilde{\mathcal R}^{D}(\kappa_0).
\eea
We can then explicitly check that all powers of $r_0$ exactly cancel between the Wronskian, $\cK^{far}_\kappa$ and $B_{lm}$. The final Teukolsky perturbation is therefore independent of $r_0$. We now substitute $\kappa_0 = \kappa_0(\frac{\ell}{\ell_*},\chi,q ; \kappa_{S^2})$ using \eqref{k0d} and $\alpha/\kappa_0=\alpha/\kappa_0(\frac{\ell}{\ell_*},\chi,q ; \kappa_{S^2})$ using \eqref{alphakappa}.

The metric perturbation is related to the curvature perturbation as  $\delta \psi_4 \rightarrow \frac{1}{2}\p^2_{\hat t} (h_+ - i h_\times)$ when $\hat r \rightarrow \infty$. Since the oscillation timescale is locked at $\hat{\omega} = m/(2M)$ at the leading order in $\lambda$, we can directly integrate for each mode $m$ to get $h_+-i h_\times = -8M^2/m^2 \delta\psi_4$. The metric perturbation at infinity is therefore given by  
\bea
h_+-i h_\times = \frac{\mu}{\hat r} \sum_{l,m} \mathcal{A}_{lm}(\frac{\ell}{\ell_*},\chi q ; \lambda, \kappa_{S^2})  S_{lm}(\theta)e^{im\hat{\phi}-i\hat{\omega} \hat u}  \label{Ampl}
\eea
where $\mu=q M$ and $\mathcal{A}_{lm}= -8 \frac{M^4}{q m^2} B_{lm}(x_*)  \cK^{far}_{\kappa}$ is independent of $M$. This is our main result. 

When the source is in NHEK instead of near-NHEK, $\ell = \ell_*$ and the result takes the form 
\bea
h_+-i h_\times = \frac{\mu}{\hat r} \sum_{l,m} \mathcal{A}_{lm}(\chi q ; \lambda, \kappa_{S^2})  S_{lm}(\theta)e^{im\hat{\phi}-i\hat{\omega} \hat u}  \label{AmplNHEK}
\eea
with $\mathcal{A}_{lm}= -8 \frac{M^4}{q m^2} B_{lm}(x_*)  \cK^{far}$ where $B_{lm}$ is defined as \eqref{Blm} with the following replacement, 
\bea
\kappa_0 \mapsto 0,\qquad \mathcal R_{lm\omega}^{\text{in}} &\mapsto &\mathcal W_{lm\Omega}^{\text{in}},\qquad \mathcal W_\kappa \mapsto \mathcal W.
\eea
It turns out convenient to express the final answer in terms of $\tilde B_{lm}(x^*)$ defined as $B_{lm}(x_*) = -\frac{qm^2}{8M^4}R_0^h \tilde B_{lm}(x^*)$. Explicitly,
\bea
\hspace{-17pt}\tilde B_{lm}(x_*)\hspace{-7pt}&=\hspace{-7pt}& -\frac{8}{(\mathcal W R_0^{2h-1})m^2} \Big\{R_0^{h-2} \mathcal W_{lm\Omega}^{\text{in}}(R_0) \Big[ \mbox{}_{-2}t_{4} \Big(V^2(R_0)+R_0^2 V''(R_0) \Big) - b_3 R_0 V'(R_0) \nn\\
&& +b_0+b_2 V(R_0)\Big] +R_0^{h-1} \mathcal W_{lm\Omega}^{\text{in}\prime}(R_0)  \Big[ -b_1 +2 \mbox{}_{-2}t_{4} R_0 V'(R_0)-\tilde b_3 V(R_0)\Big] \Big\}\label{BlmNHEK}
\eea
where $\Omega = -\alpha m R_0$ and 
\bea
b_0 &=&\mbox{}_{-2}t_{0}+ 4 (\mbox{}_{-2}t_{1}+5( \mbox{}_{-2}t_{2}+ 6(\mbox{}_{-2}t_{3} +7\mbox{}_{-2}t_{4})))  ,\nn\\
b_1 &=&\mbox{}_{-2}t_{1}+ 6\mbox{}_{-2}t_{2}+38 \mbox{}_{-2}t_{3}+ 272 \mbox{}_{-2}t_{4} ,\nn\\
b_2 &=& \mbox{}_{-2}t_{2}+12 \mbox{}_{-2}t_{3}+122 \mbox{}_{-2}t_{4},\label{bNHEK}\\
b_3 &=& \mbox{}_{-2}t_{3} + 18 \mbox{}_{-2}t_{4},     \nn   \\
\tilde b_3 &=& \mbox{}_{-2}t_{3} + 16  \mbox{}_{-2}t_{4}.  \nn   
\eea
All the coefficients $\mbox{}_{-2}t_i$ are understood to be evaluated at $\kappa_0=0$ in NHEK. As near-NHEK case, we can define $\tilde{\mathcal{W}}_{lm\Omega}^{\text{in}}$ 
\be
\mathcal{W}_{lm\Omega}^{\text{in}}(R_0)=R_0^{-h+2}\tilde{\mathcal{W}}_{lm\Omega}^{\text{in}}.
\ee 
As a result, we obtain the following scaling
\be
\tilde B_{lm}\propto R_0^{0},\qquad \mathcal A_{lm}\propto R_0^{h}.
\ee 
Since the NHEK coordinate $R$ can be mapped to far region coordinate $\hat{x}=\frac{\hat{r}-\hat{r}_+}{\hat{r}_+}$ by $\hat{x}=\lambda^{2/3}R$, 
the amplitude is proportional to $\hat{x}_0^h=(2\lambda^2)^{h/3}$ and finally reads as
\be
\mathcal{A}_{lm}(\chi q ; \lambda, \kappa_{S^2})= \frac{k_1 \hat{x}_0^h \tilde B_{lm}}{1-(i m \alpha \hat x_0)^{2h-1}k_2} .\label{scaling}
\ee

\section{Generic equatorial orbits from conformal symmetry}
\label{genericeqorbits}

So far, we only described the circular orbits on the equatorial plane of near-NHEK and NHEK. Let us denote these orbits respectively as Circular$(\ell)$  and Circular$_*$ as in \cite{Compere:2017hsi}. We showed that these orbits are the only finite size deformations of the equatorial circular orbits of a spinless particle. It turns out that the resulting spin is aligned to the orbital angular momentum. Now, all equatorial orbits can be classified under the action of the conformal group $SL(2,\mathbb R) \times U(1) \times PT$. It was shown in \cite{Compere:2017hsi} that all plunging or osculating orbits entering into the near-NHEK or NHEK region are conformally related to a circular equatorial orbit, either Circular$(\ell)$ or Circular$_*$. Such conformal transformations either preserve the NHEK or near-NHEK metric or transform one metric into the other. The same conformal maps apply to the orbits with finite size corrections. Indeed, a conformal transformation will transform covariantly the momenta and spin tensor, but since the MPD equations are covariant the resulting tensors will remain a solution to the MPD equations in either near-NHEK or NHEK. While we do not prove completeness here, we expect that any equatorial orbit can be obtained by applying one of the conformal maps listed in Appendix B.3. of \cite{Compere:2017hsi}. 

The Teukolsky waveforms for a generic equatorial orbits were obtained from the conformal map in terms of the circular orbit waveforms in \cite{Compere:2017hsi}. Finite size corrections are simply included by upgrading the physical parameters such as the energy, specific angular momentum and frequency as well as the coefficients of the radial Teukolsky equation of circular orbits to include finite size corrections. The time dependence of the waveform in the leading order high spin limit is entirely determined by the conformal class of the orbit. Finite size corrections will amount to a field redefinition of the parameters of the waveform in terms of the physical parameters (energy, angular momentum and spin).

\section{Properties of the observables}
\label{propobs}

We obtained the amplitude \eqref{Ampl} for a circular equatorial orbit in the high spin limit. Let us now discuss its properties.

\subsection{Frequency shift}

The frequency of emitted gravitational waves is locked by the kinematics to be around the extremal value
\bea
\hat \omega_{ext} = \frac{m}{2M}.
\eea 
For a circular orbit in NHEK and in the high spin limit, the frequency is given by 
\bea
\hat \omega = \frac{m}{2M}(1- \lambda^{2/3}\alpha R_0)
\eea
where $R_0$ ought to be fixed to the ISCO \eqref{ISCOspin} once the NHEK region has been glued to the asymptotically flat region. This follows from the change of coordinates between the NHEK region and the asymptotically flat region \eqref{chgt}. Using \eqref{defalpha}, the relative shift of angular frequency with respect to the extremal limit is always negative and given by
\bea
\frac{\hat{\omega}-\hat{\omega}_{\text{ext}}}{\hat{\omega}_{\text{ext}}}&=&\lambda^{2/3}(-\frac{3}{2^{5/3}}-\frac{9}{4\times 2^{2/3}}\chi q+\frac{9(-3+2\kappa_{S^2})}{4\times 2^{2/3}}(\chi q)^2+\mathcal{O}(q^3)).
\eea  
We observe that a positive secondary spin tends to slightly lower this relative shift. 

For a circular orbit in near-NHEK, $\phi=-\alpha r_0 t$ and the wave perturbation has a frequency $\omega=m\tilde{\omega}$ with  $\tilde{\omega}=-\alpha r_0$. In terms of the asymptotically flat Boyer-Linquist frame, the frequency is 
\bea
\hat \omega = \frac{m}{2M} (1- \lambda \frac{\alpha}{\kappa_0}).\label{hato}
\eea
We therefore find the relative shift of angular frequency 
\bea
\frac{\hat \omega - \hat \omega_{ext}}{\hat \omega_{ext}} = - \frac{\sqrt{3}}{2}  \frac{\lambda}{ \sqrt{1- \frac{\ell_*^2}{\ell^2}}}(1 + \frac{1}{2}(\kappa_{S^2}-1)\chi^2 q^2)+O(q^3).\label{freqs}
\eea
As already observed in \cite{Compere:2017hsi}, the description of the near-NHEK orbit displays a critical behavior in the limit $\ell \rightarrow \ell_*$, which appears in physical quantities such as the frequency shift \eqref{freqs}. One natural question is whether the enhancement factor 
\bea
\frac{1}{\sqrt{1-\frac{\ell_*^2}{\ell^2}}}
\eea
may actually lead to divergences. 

We now observe from \eqref{hato} that the near-NHEK approximation requires that $\frac{\alpha}{\kappa_0} \ll \lambda^{-1}$. Using \eqref{alphakappa}, this amounts to 
\bea
\frac{\lambda}{\sqrt{1-\frac{\ell_*^2}{\ell^2}}} \ll 1.\label{smallp}
\eea
%This is equivalent to 
%\be
%\frac{\ell_*}{| \ell |} \ll \frac{|J|}{M^2}.
%\ell \gg \ell_* .
%\ee
Therefore, the critical behavior $\ell \rightarrow \ell_*$ is never exactly reached for a given near-extremality parameter $\lambda$. Yet, for $\lambda$ very small, there is a large enhancement factor since one can reach a value close to $\ell_*$ while remaining within the near-NHEK approximation. 

The frequency shift \eqref{freqs} is maximal at the closest angular momentum $\ell$ to the critical one $\ell_*$ where the near-NHEK approximation is valid. It is minimal in the limit of large angular momentum $\ell \rightarrow \infty$. Interestingly, for a black hole with $\kappa_{S^2}=1$, the spin-induced quadrupole only contributes to the frequency shift through the shift of the critical specific angular momentum \eqref{lstar}.

\subsection{Amplitude}

As already noticed in \cite{Kesden:2011ma,Colleoni:2015afa,Gralla:2015rpa}, the leading contribution to the amplitude is $\sim \sqrt{\lambda}$ and comes from the modes $l,m$ with weight 
\bea
h = \frac{1}{2}- i \zeta_{lm}, \qquad \zeta_{lm} > 0.\label{hhalf}
\eea
In the high spin limit, only these modes are relevant. Hence, we neglect the other modes. 

For the near-NHEK circular orbit, we observe that the amplitude displays the following critical behavior
\bea
\text{lim}_{\ell \rightarrow \ell_*} \mathcal{A}_{lm}(\frac{\ell}{\ell_*},\chi q , \lambda , \kappa_{S^2}) \sim \left( \frac{\lambda}{\sqrt{1-\frac{\ell_*^2}{\ell^2}}} \right)^{1/2}.
\eea
This extends the observation of \cite{Compere:2017hsi} to particles with spin and quadrupole. Now, we showed in the previous section the existence of the small parameter \eqref{smallp} for the near-NHEK approximation to be valid. The interpretation of this critical behavior is therefore that there is an amplitude enhancement with a power law behavior $\sim (1-\frac{\ell_*^2}{\ell^2})^{-1/4}$ that partially compensate the redshift of the amplitude $\sim \sqrt{\lambda}$ due to the existence of the NHEK region. The final amplitude is always finite. In that sense, the critical behavior is capped due to the matching with the asymptotically flat region. This is in essence the physics of a capped $AdS_2$ region. 

For $\lambda$ small but finite, one could use either the NHEK or the near-NHEK description of the amplitude and there should therefore be a map between the two descriptions. The near-NHEK critical behavior can be understood in terms of the NHEK amplitude as follows.  The change of coordinate between the Boyer-Linquist coordinate $\hat x = \frac{\hat r -\hat r_+}{\hat r_+}$ and near-NHEK coordinate $r$ \eqref{changecoord} implies that
\be
\hat{x}_0= \frac{\lambda}{\kappa_0} = \lambda(\frac{2\ell}{\sqrt{3(\ell^2 - \ell_*^2)}} (1+\frac{\chi q}{2}+ (1-\frac{\kappa_{S^2}}{2})\chi^2 q^2 + O(q^3))-1)
\ee 
where we used \eqref{k0d}. Therefore, we can rewrite the scaling behavior in terms of the exterior quantity $\hat x_0$ as
 \be
 \text{lim}_{\ell \rightarrow \ell_*} \mathcal{A}_{lm}(\frac{\ell}{\ell_*},\chi q , \lambda , \kappa_{S^2}) \sim \hat{x}_0^{1/2}.\label{scaling2}
 \ee  
We now see the agreement with the NHEK amplitude (\ref{scaling}) valid at $\ell = \ell_*$ after taking into account $\hat{x}_0\ll1$ and $Re(h) =\frac{1}{2}$ for the dominant contribution at high spin.  

The ringdown has particular features in the high spin regime. The spectrum of quasi-normal modes splits in the near-extremal limit into damped modes in the asymptotically flat region, and into zero-damped modes in the near-horizon region \cite{Yang:2012pj}. The presence of zero-damped modes in the near-horizon region leads to polynomial quasi-normal mode ringing due to harmonic stacking of overtones \cite{Yang:2013uba}. As discussed in \cite{Compere:2017hsi}, this polynomial ringing gets emitted for geodesic plunges and leads to a ``smoking gun'' signature of the gravitational wave emission from a plunging source into a high spin black hole (see also \cite{Gralla:2018xzo}). More precisely, for a face-on collision, only the $m=l=2$ harmonic contributes, while for an edge-on collision all harmonics $m,l$ that have a real part of the conformal weight equal to a half significantly contribute with a contribution that decays with $l$. These features are identical after finite size corrections, since the angular dependence of the waveform remains unchanged. 

\subsection{Energy fluxes}

The energy fluxes at the ISCO are crucial to determine the leading order orbit of the infalling compact object in the central black hole. Since the ISCO is in NHEK spacetime, we now specialize to this case only. The ISCO, or, in our notation, the Circular$_*$ orbit is parametrized by the spin $\chi$, the mass ratio $q$ and the radius $R_0$.  Before the spin corrections, the ISCO is located at $R_0=2^{1/3}$ as one can check using the matching with the asymptotically flat region. The ISCO is affected by the spin and quadrupole corrections. Now, we have shown that $\tilde{B}_{lm}$ is independent of $R_0$, and it is also the case of $\mathcal{K}^{\text{far}}$ since it is a quantity independent of the source. The physical quantities will therefore be independent on $R_0$ and we will set for definiteness $R_0=2^{1/3}$ in numerical computations.

The fluxes at infinity and at the horizon are given in the high spin approximation by \cite{1974ApJ...193..443T,Compere:2017hsi}
\bea
\left( \frac{dE}{d\hat t}\right)^\infty &=& \sum_{l,m} \frac{2}{m^2} |M^4 B_{lm} \mathcal K^{far}|^2 = q^2 \sum_{l,m} \frac{m^2}{32} |\tilde  B_{lm} R_0^h \mathcal K^{far}|^2\label{dEinfdt} ; \\
\left( \frac{dE}{d\hat t} \right)^H &=& -\lambda^{2/3} \sum_{l,m} \frac{2 |m| \, e^{-\pi |m|-\pi \, \text{sign}(m) \zeta_{lm}}}{ \prod_{k=0}^3 (m^2 + (h-2+k)^2)}|M^4 (C_{lm}+A_{lm})|^2\label{dEHdt}
\eea
where the summation is only for modes $l,m$ with $\text{Re}(h)=\frac{1}{2}$, see \eqref{hhalf}. In particular, $m \neq 0$. The radial solution is written in \eqref{dif} with $X_1 \mapsto C_{lm}$, $X_2 \mapsto B_{lm}$ and $Y_1 \mapsto A_{lm}$. Here, $A_{lm}$ is given in terms of $B_{lm}$ in \eqref{Alm} and $C_{lm}$ is written as $B_{lm}$ with $\mathcal W^{\text{in}}$ exchanged with $\mathcal M^{\text{D}}$ at $\mathcal W$ fixed, namely 
\bea
\hspace{-17pt} C_{lm}\hspace{-7pt}&=\hspace{-7pt}& \frac{q}{\mathcal W M^4 R_0} \Big\{ \mathcal M_{lm\Omega}^{\text{D}}(R_0) \Big[ \mbox{}_{-2}t_{4} \Big(V^2(R_0)+R_0^2 V''(R_0) \Big) - b_3 R_0 V'(R_0) \nn\\
&& +b_0+b_2 V(R_0)\Big] +R_0 \mathcal M_{lm\Omega}^{\text{D}\prime}(R_0)  \Big[ -b_1 +2 \mbox{}_{-2}t_{4} R_0 V'(R_0)-\tilde b_3 V(R_0)\Big] \Big\} . \label{ClmNHEK}
\eea
Here, the $b_i$ coefficients are given in \eqref{bNHEK}. In the absence of finite size effects, we explicitly checked that the formulae \eqref{dEinfdt}-\eqref{dEHdt} exactly match with Eq. (76)-(77) of \cite{Gralla:2015rpa}. The sign in front of \eqref{dEHdt} reflects that all modes $m \neq 0$ are superradiant. 

After a numerical evaluation taking into account all harmonics from $l=2$ up to $l=50$, we obtain the following fluxes
\bea
\left( \frac{dE}{d\hat t}\right)^\infty &=& q^2 \hat x_0 \left( a^\infty_{(0)} + a^\infty_{(1)}\chi q+ (a^\infty_{(2)} +\tilde a^\infty_{(2)} \kappa_{S^2}) (\chi q)^2+O(q^3) \right); \label{fluxinfinity}\\
\left( \frac{dE}{d\hat t} \right)^H &=& q^2 \hat x_0 \left( a^H_{(0)} + a^H_{(1)}\chi q+ (a^H_{(2)} +\tilde a^H_{(2)} \kappa_{S^2}) (\chi q)^2+O(q^3) \right)\label{fluxhorizon}
\eea
where $\hat x_0 = (2\lambda^2)^{1/3}$ and
\bea
&a^\infty_{(0)} = 0.987 \; ;\qquad & a^H_{(0)} = -0.13285\; ;\\
&a^\infty_{(1)} =-0.409 \; ;\qquad & a^H_{(1)} =0.28780 \; ;\\
&a^\infty_{(2)} = 0.784\; ;\qquad & a^H_{(2)} =-0.03169\; ;\\
&\tilde a^\infty_{(2)} = 2.889\; ;\qquad  & \tilde a^H_{(2)} =-0.70616 \; .\label{acoefficients}
\eea
Additional accuracy is possible for the horizon coefficients thanks to exponential convergence for large $m$. We numerically find that the coefficients do not depend upon $\lambda$ within numerical error. The $\lambda$ dependence is highly subleading as already noticed in Section 5 of \cite{Compere:2017hsi}. The first two coefficients $a^\infty_{(0)}$ and $a^H_{(0)}$ both agree with \cite{Gralla:2016qfw}. The two coefficients $a^\infty_{(1)}$ and $a^H_{(1)}$ characterize the leading order spin corrections while $a^\infty_{(2)}$ and $a^H_{(2)}$ characterize the subleading order spin corrections. The remaining two coefficients $\tilde{a}^\infty_{(2)}$ and $\tilde{a}^H_{(2)}$ characterize the leading order finite size effect from the quadrupole.  These flux expressions allow to derive the leading order of the near-horizon equatorial inspiral including finite size effects, which straightforwardly generalizes the point-particle equatorial inspiral of \cite{Gralla:2016qfw}. 

We will further comment on the structure of \eqref{fluxinfinity}-\eqref{fluxhorizon} and its arbitrary finite size corrections at leading order in the high spin regime $\lambda \ll 1$. For a nearly extremal black hole, the spin of the central massive black hole is determined in terms of the mass as $J=M^2$ up to $\lambda$ corrections. The mass $M$ gives the overall scale of any result. The probe mass and spin can be expressed in terms of the adimensional quantities 
 $q$ and $\chi$. Note that $q$ is a small parameter while $-1 \leq \chi \leq 1$. For a probe black hole, all multipole moments are determined by the mass and spin or, equivalently, by $q,\chi$. The $2^N$-multipole arises exactly at order $q^N \chi^N$ with respect to the particule leading order result since it is proportional to the spin to the $N$-th power and since it is suppressed by $q^N$. The energy fluxes at the horizon or at infinity sourced by a black hole including all finite size corrections (but no self-force effects) therefore take the form
\be
\left( \frac{dE}{d\hat t}\right)^{\infty,H}=q^2\hat{x}_0\mathcal{F}^{\infty,H}(\chi q)
\ee 
at the leading order in $\lambda$.

\section{Conclusion}

We derived the Teukolsky perturbation at leading order in the high spin regime of a finite size compact object orbiting a circular equatorial orbit in the near-horizon region of a highly spinning massive black hole. This extends the result derived 
 in NHEK \cite{Gralla:2015rpa} and in near-NHEK \cite{Compere:2017hsi} for a point particle to a spinning particle with spin induced quadrupole.  We determined along the way how to recover NHEK results in terms of a limit from the near-NHEK results. We also indicated how the Teukolsky perturbation can be obtained for plunging equatorial orbits using conformal $SL(2,\mathbb R) \times U(1) \times PT$ transformations following the method of \cite{Hadar:2014dpa} and its classification \cite{Compere:2017hsi}. We discussed the spin and quadrupole corrections to the frequency of emission and to the amplitude. We obtained the flux formulae at the horizon and at infinity, which generalizes the fluxes obtained in \cite{Gralla:2015rpa,Gralla:2016qfw} to include spin and quadrupole couplings. Our results cannot be directly compared with post-Newtonian results since we use the strong-field (near-)NHEK spacetime as background geometry of the ISCO orbit.

Finite size effects allow to distinguish black holes from neutron stars in precise gravitational wave observations. In particular, the measurement of $\kappa_{S^2}$, the amplitude of the spin induced quadrupole, encodes information about the internal structure of neutron stars. Theoretically, black holes are very special as we have noticed in (\ref{defe}), (\ref{pt}), (\ref{pphi}) and (\ref{freqs}). The spin induced quadrupole exactly cancels the subleading order spin effect in these expressions. We expect that similar cancellations will occur at subleading orders. This motivates us to solve in the future the MPD equations for black hole coalescences including the entire spin-induced multipole tower. 

 Our study focused on binary systems with a central compact object being of extremely high spin. The semi-analytic results presented in this work allow for new consistency tests of numerical codes in the high spin limit.
Our analytic results need to be further extended in several directions to achieve the goal of providing accurate modeling of intermediate mass-ratio coalescences. First, the first and second order self-force needs to be included, which is a formidable task. Second, the quasi-circular and equatorial hypotheses need to be relaxed in order to model eccentric and inclined orbits (for recent work in that direction see e.g. \cite{vandeMeent:2015lxa,vandeMeent:2016pee,vandeMeent:2017bcc}). Finally, the leading high spin limit needs to be extended to include perturbative corrections away from maximal spin  while preserving a high level of analytic control. The result of this paper is only one milestone in this research program, which will be continued elsewhere. 

\section*{Acknowledgments} 

We are grateful to Kwinten Fransen, Adam Pound, Maarten van de Meent and Niels Warburton for useful discussions.  We would like to thank Peng-cheng Li for his participation at the early stage of the project. This work was supported in part by the European Research Council Starting Grant 335146 ``HoloBHC", NSFC Grants No.~11275010, No.~11335012, No.~11325522 and No. 11735001. This work makes use of the Black Hole Perturbation Toolkit. We also acknowledge networking support by the COST Action GWverse CA16104. G.C. is a Research Associate of the Fonds de la Recherche Scientifique F.R.S.-FNRS (Belgium). The research of J.L. was supported by the Ministry of Science, ICT \& Future Planning, Gyeongsangbuk-do and Pohang City. Y.L. is also financially supported by the China Scholarship Council.

\appendix

\section{Coefficients of the radial source in near-NHEK}
\label{coefst4t0}

The $\mbox{}_{-2}t_{i}$ and $\mbox{}_{+2}t_{i}$ coefficients, $i=0,\dots 4$, determining the source of the radial Teukolsky equation corresponding to the circular equatorial orbit in near-NHEK in the extreme mass ratio limit and including spin and quadrupole couplings are given by
\bea
\mbox{}_{s}t_{i} = \mbox{}_{s}t^{(0)}_{i} +  \mbox{}_{s}t^{(1)}_{i} \chi q+   \mbox{}_{s}t^{(2)}_{i} (\chi q)^2 + O(q^3)
\eea
where the coefficients are
\bea
\mbox{}_{-2}t^{(0)}_3\hspace{-6pt}&=&\hspace{-6pt}\mbox{}_{-2}t^{(0)}_4=\mbox{}_{-2}t^{(1)}_4=\mbox{}_{+2}t^{(0)}_3=\mbox{}_{+2}t^{(0)}_4=\mbox{}_{+2}t^{(1)}_4=0;
\eea
\bea
\mbox{}_{-2}t^{(0)}_0\hspace{-6pt}&=&\hspace{-6pt}-\frac{1}{128\lambda_0}[(-m^2 (-3 - 6 \kappa_0 + \kappa_0^2)^2 +64 (5 + 20 \kappa_0 + 25 \kappa_0^2 + 10 \kappa_0^3 +2 \kappa_0^4)\nn\\
&& -8 i m (-3 - 12 \kappa_0 - 20 \kappa_0^2 - 16 \kappa_0^3 +3 \kappa_0^4))S_{lm}(\frac{\pi}{2})+(-128 i + 48 m - 512 i \kappa_0 \nn\\
&&+ 192 m \kappa_0-768 i \kappa_0^2 + 176 m \kappa_0^2 - 512 i \kappa_0^3 -32 m \kappa_0^3)S'_{lm}(\frac{\pi}{2})-64 (1 + 2 \kappa_0)^2S''_{lm}(\frac{\pi}{2})];\nn\\
\mbox{}_{-2}t^{(1)}_0\hspace{-6pt}&=&\hspace{-6pt}-\frac{1}{256\lambda_0^3}[(-i m^3 (-3 - 6 \kappa_0 + \kappa_0^2)^3 +m^2 (-3 - 6 \kappa_0 + \kappa_0^2)^2 (-11 - 22 \kappa_0 + 29 \kappa_0^2)\nn\\
&& +16 i m \kappa_0^2 (39 + 156 \kappa_0 + 110 \kappa_0^2 -92 \kappa_0^3 + 11 \kappa_0^4) +64 (-123 - 738 \kappa_0 - 1514 \kappa_0^2 \nn\\
&&- 1136 \kappa_0^3 -81 \kappa_0^4 + 142 \kappa_0^5 + 2 \kappa_0^6))S_{lm}(\frac{\pi}{2})+2 i (m^2 (-7 - 14 \kappa_0 + \kappa_0^2) (-3 -6 \kappa_0 \nn\\
&&+ \kappa_0^2)^2 +24 i m (-6 - 36 \kappa_0 - 55 \kappa_0^2 + 20 \kappa_0^3 +60 \kappa_0^4 - 16 \kappa_0^5 + \kappa_0^6) -64 (-30 \nn\\
&&- 180 \kappa_0 - 361 \kappa_0^2 - 244 \kappa_0^3 +3 \kappa_0^4 + 14 \kappa_0^5 + 2 \kappa_0^6))S'_{lm}(\frac{\pi}{2})+16 i (1 + 2 \kappa_0) (m (-3 \nn\\
&&- 6 \kappa_0 + \kappa_0^2)^2 +4 i (-3 - 12 \kappa_0 - 19 \kappa_0^2 - 14 \kappa_0^3 +4 \kappa_0^4))S''_{lm}(\frac{\pi}{2})];\nn
\eea
\bea
\mbox{}_{-2}t^{(2)}_0\hspace{-6pt}&=&\hspace{-6pt}\frac{1}{4096(1+2\kappa_0)\lambda_0^5}[(\kappa_{S^2}(-3 - 6 \kappa_0 + \kappa_0^2) (m^4 (-3 -6 \kappa_0 + \kappa_0^2)^4 +24i m^3 (-3 - 6 \kappa_0 \nn\\
&&+ \kappa_0^2)^3 (-1 -2 \kappa_0 + \kappa_0^2) -1536 i m (15 + 120 \kappa_0 + 361 \kappa_0^2 + 486 \kappa_0^3 +247 \kappa_0^4 - 12\kappa_0^5\nn\\
&&- 19 \kappa_0^6 +2 \kappa_0^7) -2048 (-183 - 1464 \kappa_0 - 4480 \kappa_0^2 -6384 \kappa_0^3 - 3927 \kappa_0^4 - 476 \kappa_0^5 \nn\\
&&+234 \kappa_0^6 + 12 \kappa_0^7) -64 m^2 (9 + 72 \kappa_0 + 186 \kappa_0^2 + 108 \kappa_0^3 -189 \kappa_0^4 - 132 \kappa_0^5 + 96 \kappa_0^6\nn\\
&&-24 \kappa_0^7 + 2 \kappa_0^8))-24 i (1 + \kappa_0)^2 (1 +2 \kappa_0) (3 m^3 (-3 - 6 \kappa_0 + \kappa_0^2)^3 +2 i m^2 (-3 - 6 \kappa_0\nn\\
&&+ \kappa_0^2)^2 (-53 - 106 \kappa_0 +30 \kappa_0^2) -64 i (-339 - 2034 \kappa_0 - 3874 \kappa_0^2 -1936 \kappa_0^3 + 819 \kappa_0^4 \nn\\
&&+ 86 \kappa_0^5 +2 \kappa_0^6) -8 m (-585 - 3510 \kappa_0 - 6291 \kappa_0^2 - 1764 \kappa_0^3 +2645 \kappa_0^4 - 542 \kappa_0^5 \nn\\
&&+ 31 \kappa_0^6)))S_{lm}(\frac{\pi}{2})+(16\kappa_{S^2} (1 + 2 \kappa_0) (-3 -6 \kappa_0 + \kappa_0^2) (m^3 (-3 -6 \kappa_0 + \kappa_0^2)^3 \nn\\
&&+20 i m^2 (-3 - 6 \kappa_0 + \kappa_0^2)^2 (-1 -2 \kappa_0 + \kappa_0^2) -32 m \kappa_0^2 (15 + 60 \kappa_0 + 52 \kappa_0^2 -16 \kappa_0^3 \nn\\
&&+ \kappa_0^4) +256 i (-42 - 252 \kappa_0 - 527 \kappa_0^2 - 428 \kappa_0^3 -81 \kappa_0^4 + 22 \kappa_0^5 + 2 \kappa_0^6))+48 i (1 \nn\\
&&+ \kappa_0)^2 (1 +2 \kappa_0) (m^2 (-3 - 6 \kappa_0 + \kappa_0^2)^2 (-13 - 26 \kappa_0 + 3 \kappa_0^2)-64 (57 + 342 \kappa_0 + 680 \kappa_0^2\nn\\
&& + 440 \kappa_0^3 -41 \kappa_0^4 - 50 \kappa_0^5 + 2 \kappa_0^6) +24 i m (-33 - 198 \kappa_0 - 358 \kappa_0^2 - 112 \kappa_0^3 +137 \kappa_0^4 \nn\\
&&- 30 \kappa_0^5 + 2 \kappa_0^6)))S'_{lm}(\frac{\pi}{2})+(64\kappa_{S^2} (1 + 2 \kappa_0)^2 (-3 -6 \kappa_0 + \kappa_0^2) (8 i m (-3 -6 \kappa_0 \nn\\
&&+ \kappa_0^2)^2 +m^2 (-3 - 6 \kappa_0 + \kappa_0^2)^2 + 32 (-15 - 60 \kappa_0 - 65 \kappa_0^2 - 10 \kappa_0^3 +4 \kappa_0^4)) +384 i (1 \nn\\
&&+ \kappa_0)^2 (1 +2 \kappa_0)^2 (m (-3 - 6 \kappa_0 + \kappa_0^2)^2 -8 i (-15 - 60 \kappa_0 - 52 \kappa_0^2 + 16 \kappa_0^3)))S''_{lm}(\frac{\pi}{2})];\nn\\
\mbox{}_{-2}t^{(0)}_1\hspace{-6pt}&=&\hspace{-6pt}-\frac{i(1+\kappa_0)(1+2\kappa_0)}{16\lambda_0}[(16 i (1 + \kappa_0)^2 + m (-3 - 6 \kappa_0 + \kappa_0^2))S_{lm}(\frac{\pi}{2})\nn\\
&&+8 (1 + 2 \kappa_0)S'_{lm}(\frac{\pi}{2})];\nn\\
\mbox{}_{-2}t^{(1)}_1\hspace{-6pt}&=&\hspace{-6pt}\frac{(1+\kappa_0)(1+2\kappa_0)}{64\lambda_0^3}[(m^2 (-3 - 6 \kappa_0 + \kappa_0^2)^2 -4 i m (27 + 108 \kappa_0 + 96 \kappa_0^2 -24 \kappa_0^3 + \kappa_0^4) \nn\\
&&+32 (-72 - 288 \kappa_0 - 287 \kappa_0^2 + 2 \kappa_0^3 +13 \kappa_0^4))S_{lm}(\frac{\pi}{2})-4 (m (-3 - 6 \kappa_0 + \kappa_0^2)^2 \nn\\
&&+16 i (-12 - 48 \kappa_0 - 46 \kappa_0^2 +4 \kappa_0^3 + \kappa_0^4))S'_{lm}(\frac{\pi}{2})+16 (3 + 12 \kappa_0 + 11 \kappa_0^2\nn\\
&& - 2 \kappa_0^3)S''_{lm}(\frac{\pi}{2})];\nn
\eea
\bea
\mbox{}_{-2}t^{(2)}_1\hspace{-6pt}&=&\hspace{-6pt}\frac{(1+\kappa_0)}{512\lambda_0^5}[(\kappa_{S^2}(-3 - 6 \kappa_0 + \kappa_0^2) (-i m^3 (-3 -6 \kappa_0 + \kappa_0^2)^3 +12 m^2 (3 + 9 \kappa_0 + 5 \kappa_0^2 \nn\\
&&- \kappa_0^3)^2 +768 (-77 - 462 \kappa_0 - 944 \kappa_0^2 - 696 \kappa_0^3 -69 \kappa_0^4 + 22 \kappa_0^5 + 2 \kappa_0^6) -32 i m (90 \nn\\
&&+ 540 \kappa_0 + 1059 \kappa_0^2 + 636 \kappa_0^3 -102 \kappa_0^4 - 36 \kappa_0^5 + 5 \kappa_0^6))-24 (1 + 2 \kappa_0) (m^2 (-3 - 6 \kappa_0 \nn\\
&&+ \kappa_0^2)^2 (-1 -2 \kappa_0 + \kappa_0^2) +i m (-297 - 1782 \kappa_0 - 3333 \kappa_0^2 - 1452 \kappa_0^3 +877 \kappa_0^4 - 94 \kappa_0^5 \nn\\
&&+ \kappa_0^6) -16 (-225 - 1350 \kappa_0 - 2745 \kappa_0^2 - 1980 \kappa_0^3 -111 \kappa_0^4 + 138 \kappa_0^5 + \kappa_0^6)))S_{lm}(\frac{\pi}{2})\nn\\
&&+(-8 i\kappa_{S^2} (1 + 2 \kappa_0) (-3 -6 \kappa_0 + \kappa_0^2) (-16 i m (-3 -6 \kappa_0 + \kappa_0^2)^2 +m^2 (-3 - 6 \kappa_0 \nn\\
&&+ \kappa_0^2)^2 +32 (-87 - 348 \kappa_0 - 365 \kappa_0^2 - 34 \kappa_0^3 +16 \kappa_0^4))+96 (1 + 2 \kappa_0) (m (-3 - 6 \kappa_0 \nn\\
&&+ \kappa_0^2)^2 (-1 -2 \kappa_0 + \kappa_0^2) +4 i (51 + 306 \kappa_0 + 643 \kappa_0^2 + 532 \kappa_0^3 +98 \kappa_0^4 \nn\\
&&- 52 \kappa_0^5+2\kappa_0^6)))S'_{lm}(\frac{\pi}{2})+(-1024\kappa_{S^2} (1 + 2 \kappa_0)^2 (-3 - 6 \kappa_0 + \kappa_0^2)^2\nn\\
&&-768 (1 + 2 \kappa_0)^3 (-3 - 6 \kappa_0 + \kappa_0^2))S''_{lm}(\frac{\pi}{2})];\nn\\
\mbox{}_{-2}t^{(0)}_2\hspace{-6pt}&=&\hspace{-6pt}-\frac{(1+\kappa_0)^2(1+2\kappa_0)^2S_{lm}(\frac{\pi}{2})}{8\lambda_0};\nn\\
\mbox{}_{-2}t^{(1)}_2\hspace{-6pt}&=&\hspace{-6pt}\frac{i(1+2\kappa_0)^2}{16\lambda_0^3}[(m (-3 - 6 \kappa_0 + \kappa_0^2)^2 +i (-159 - 636 \kappa_0 - 714 \kappa_0^2 - 156 \kappa_0^3 \nn\\
&&+49 \kappa_0^4)) S_{lm}(\frac{\pi}{2}) +2 (-15 - 60 \kappa_0 - 58 \kappa_0^2 + 4 \kappa_0^3 + \kappa_0^4)S'_{lm}(\frac{\pi}{2})];\nn\\
\mbox{}_{-2}t^{(2)}_2\hspace{-6pt}&=&\hspace{-6pt}-\frac{(1+2\kappa_0)}{64\lambda_0^5}[(\kappa_{S^2}(-3 - 6 \kappa_0 + \kappa_0^2) (m^2 (1 + 2 \kappa_0) (-3 -6 \kappa_0 + \kappa_0^2)^2\nn\\
&&-2 i m (-3 - 6 \kappa_0 + \kappa_0^2)^2 (11 + 22 \kappa_0 +9 \kappa_0^2) +48 (-53 - 318 \kappa_0 - 683 \kappa_0^2 - 612 \kappa_0^3 \nn\\
&&-184 \kappa_0^4 + 8 \kappa_0^5 + 6 \kappa_0^6)) +6 i (1 + \kappa_0)^2 (1 +2 \kappa_0) (m (-3 - 6 \kappa_0 + \kappa_0^2)^2 -i (-573 \nn\\
&&- 2292 \kappa_0 - 2080 \kappa_0^2 +424 \kappa_0^3 + \kappa_0^4)))S_{lm}(\frac{\pi}{2})+(-4 \kappa_{S^2}(-3 - 6 \kappa_0 + \kappa_0^2) (-3 - 12 \kappa_0 \nn\\
&&-11 \kappa_0^2 +2 \kappa_0^3)(m (-3 - 6 \kappa_0 + \kappa_0^2) +4 i (15 + 30 \kappa_0 + 13 \kappa_0^2))+12 i (1 + \kappa_0)^2 (-11 \nn\\
&&- 22 \kappa_0 + \kappa_0^2) (-3 -12 \kappa_0 - 11 \kappa_0^2 + 2 \kappa_0^3))S'_{lm}(\frac{\pi}{2})\nn\\
&&-16\kappa_{S^2} ((3 + 12 \kappa_0 + 11 \kappa_0^2 - 2 \kappa_0^3)^2) S''_{lm}(\frac{\pi}{2})];\nn\\
\mbox{}_{-2}t^{(1)}_3\hspace{-6pt}&=&\hspace{-6pt}-\frac{(1+\kappa_0)(1+2\kappa_0)^3}{4\lambda_0}S_{lm}(\frac{\pi}{2});\nn\\
\mbox{}_{-2}t^{(2)}_3\hspace{-6pt}&=&\hspace{-6pt}\frac{(1+\kappa_0)(1+2\kappa_0)^2}{32\lambda_0^3}[(i\kappa_{S^2} (-3 - 6 \kappa_0 + \kappa_0^2) (m (-3 -	6 \kappa_0 + \kappa_0^2) +	4 i (13 + 26 \kappa_0 \nn\\
&&+ 9 \kappa_0^2))-48 (1 + 2 \kappa_0)^2)S_{lm}(\frac{\pi}{2})-8 i\kappa_{S^2} (3 + 12 \kappa_0 + 11 \kappa_0^2 - 2 \kappa_0^3)S'_{lm}(\frac{\pi}{2})];\nn\\
\mbox{}_{-2}t^{(2)}_4\hspace{-6pt}&=&\hspace{-6pt}-\frac{(1+\kappa_0)^2(1+2\kappa_0)^3\kappa_{S^2}S_{lm}(\frac{\pi}{2})}{16\lambda_0} ;\nn
\eea
\bea
\mbox{}_{+2}t^{(0)}_0\hspace{-6pt}&=&\hspace{-6pt}-\frac{1}{32(1+2\kappa_0)^2\lambda_0}[(128 (1 + 2 \kappa_0)^2 - m^2 (-3 - 6 \kappa_0 + \kappa_0^2)^2 -8 i m (-9 - 36 \kappa_0 \nn\\
&&- 36 \kappa_0^2 + \kappa_0^4))S_{lm}(\frac{\pi}{2})+16 (1 + 2 \kappa_0) (8 i (1 + 2 \kappa_0) +m (-3 - 6 \kappa_0 + \kappa_0^2))S'_{lm}(\frac{\pi}{2})\nn\\
&&-64 (1 + 2 \kappa_0)^2S''_{lm}(\frac{\pi}{2})];\nn\\
\mbox{}_{+2}t^{(1)}_0\hspace{-6pt}&=&\hspace{-6pt}-\frac{i}{64(1+2\kappa_0)^2\lambda_0^3}[(m^3 (-3 - 6 \kappa_0 + \kappa_0^2)^3 -128 i (1 + 2 \kappa_0)^2 (9 + 18 \kappa_0 + \kappa_0^2)\nn\\
&& -i m^2 (-3 - 6 \kappa_0 + \kappa_0^2)^2 (-27 - 54 \kappa_0 +13 \kappa_0^2) +16 m (72 + 432 \kappa_0 + 861 \kappa_0^2 + 564 \kappa_0^3 \nn\\
&&-46 \kappa_0^4 - 68 \kappa_0^5 + 9 \kappa_0^6))S_{lm}(\frac{\pi}{2})+2 (m^2 (-7 - 14 \kappa_0 + \kappa_0^2) (-3 -6 \kappa_0 + \kappa_0^2)^2 \nn\\
&&+64 (1 + 2 \kappa_0)^2 (9 + 18 \kappa_0 + \kappa_0^2) +8 i m (54 + 324 \kappa_0 + 609 \kappa_0^2 + 276 \kappa_0^3 -152 \kappa_0^4 \nn\\
&&+ 8 \kappa_0^5 + \kappa_0^6))S'_{lm}(\frac{\pi}{2})-16 (1 + 2 \kappa_0) (m (-3 - 6 \kappa_0 + \kappa_0^2)^2-4i(9 + 36\kappa_0\nn\\
&&+37 \kappa_0^2 + 2 \kappa_0^3)) S''_{lm}(\frac{\pi}{2})];\nn\\
\mbox{}_{+2}t^{(2)}_0\hspace{-6pt}&=&\hspace{-6pt}\frac{1}{1024(1+2\kappa_0)^3\lambda_0^5}[(\kappa_{S^2}(-3 - 6 \kappa_0 + \kappa_0^2) (m^4 (-3 -6 \kappa_0 + \kappa_0^2)^4 -4096 (1 + 2 \kappa_0)^3 (3 \nn\\
&&+ 6 \kappa_0 + \kappa_0^2) +8 i m^3 (-3 - 6 \kappa_0 + \kappa_0^2)^3 (7 +14 \kappa_0 + \kappa_0^2) +1024 i m \kappa_0^2 (-3 - 18 \kappa_0 \nn\\
&&- 23 \kappa_0^2 +28 \kappa_0^3 + 42 \kappa_0^4 - 20 \kappa_0^5 +2 \kappa_0^6) +64 m^2 (-135 - 1080 \kappa_0 - 3234 \kappa_0^2 -4284 \kappa_0^3 \nn\\
&&- 2003 \kappa_0^4 + 388 \kappa_0^5 +304 \kappa_0^6 - 72 \kappa_0^7 + 4 \kappa_0^8))+24 i (1 + \kappa_0)^2 (1 +2 \kappa_0) (2304 i (1 \nn\\
&&+ 2 \kappa_0)^3 +3 m^3 (-3 - 6 \kappa_0 + \kappa_0^2)^3 -2 i m^2 (-3 - 6 \kappa_0 + \kappa_0^2)^2 (-37 - 74 \kappa_0 +14 \kappa_0^2)\nn\\
&& +24 m (-81 - 486 \kappa_0 - 843 \kappa_0^2 - 132 \kappa_0^3 +461 \kappa_0^4 - 110 \kappa_0^5 + 7 \kappa_0^6)))S_{lm}(\frac{\pi}{2})\nn\\
&&+(-16\kappa_{S^2} (1 + 2 \kappa_0) (-3 -6 \kappa_0 + \kappa_0^2) (-4 i m^2 (-9 -18 \kappa_0 + \kappa_0^2) (-3 - 6 \kappa_0 + \kappa_0^2)^2 \nn\\
&&+m^3 (-3 - 6 \kappa_0 + \kappa_0^2)^3 +256 i (1 + 2 \kappa_0)^2 (3 + 6 \kappa_0 + \kappa_0^2) +32 m (18 + 108 \kappa_0 \nn\\
&&+ 225 \kappa_0^2 + 180 \kappa_0^3 +22 \kappa_0^4 - 28 \kappa_0^5 + 3 \kappa_0^6))+48 i (1 + \kappa_0)^2 (1 + 2 \kappa_0) (-1152 (1 \nn\\
&&+ 2 \kappa_0)^3 +m^2 (-3 - 6 \kappa_0 + \kappa_0^2)^2 (-13 - 26 \kappa_0 +3 \kappa_0^2) +8 i m (27 + 162 \kappa_0 + 330 \kappa_0^2 \nn\\
&&+ 240 \kappa_0^3 +13 \kappa_0^4 - 22 \kappa_0^5 + 2 \kappa_0^6)))S'_{lm}(\frac{\pi}{2})+(64\kappa_{S^2} (1 + 2 \kappa_0)^2 (-3 -6 \kappa_0 \nn\\
&&+ \kappa_0^2) (-8 i m (-3 -6 \kappa_0 + \kappa_0^2)^2 +m^2 (-3 - 6 \kappa_0 + \kappa_0^2)^2 +32 (3 + 12 \kappa_0 \nn\\
&&+ 13 \kappa_0^2 + 2 \kappa_0^3))-384 i (1 + \kappa_0)^2 (1 +2 \kappa_0)^2 (72 i (1 + 2 \kappa_0)^2 +m (-3 - 6 \kappa_0 \nn\\
&&+ \kappa_0^2)^2))S''_{lm}(\frac{\pi}{2})] ;\nn\\
\mbox{}_{+2}t^{(0)}_1\hspace{-6pt}&=&\hspace{-6pt}\frac{i(1+\kappa_0)}{4(1+2\kappa_0)\lambda_0}[m(-3-6\kappa_0+\kappa_0^2)S_{lm}(\frac{\pi}{2})
-8(1+2\kappa_0)S'_{lm}(\frac{\pi}{2})];\nn
\eea
\bea
\mbox{}_{+2}t^{(1)}_1\hspace{-6pt}&=&\hspace{-6pt}\frac{(1+\kappa_0)}{16(1+2\kappa_0)\lambda_0^3}[(-3 - 6 \kappa_0+ \kappa_0^2) (32 + 64 \kappa_0 +m^2 (-3 - 6 \kappa_0 + \kappa_0^2) \nn\\
&&-4 i m (-15 - 30 \kappa_0 + 7 \kappa_0^2))S_{lm}(\frac{\pi}{2})+4 (-48 i (1 + 2 \kappa_0)^2 + m (-3 - 6 \kappa_0 \nn\\
&&+ \kappa_0^2)^2)S'_{lm}(\frac{\pi}{2})+16 (3 + 12 \kappa_0 + 11 \kappa_0^2 - 2 \kappa_0^3)S''_{lm}(\frac{\pi}{2})];\nn\\
\mbox{}_{+2}t^{(2)}_1\hspace{-6pt}&=&\hspace{-6pt}\frac{(1+\kappa_0)}{128(1+2\kappa_0)^2\lambda_0^5}[(\kappa_{S^2}(i m (-3 - 6 \kappa_0 + \kappa_0^2)^2 (m^2 (-3 -6 \kappa_0 + \kappa_0^2)^2 -4 i m (39 + 156 \kappa_0 \nn\\
&&+ 134 \kappa_0^2 - 44 \kappa_0^3 +3 \kappa_0^4) +32 (12 + 48 \kappa_0 + 31 \kappa_0^2 - 34 \kappa_0^3 +5 \kappa_0^4)))-24 (1 + 2 \kappa_0)(-3\nn\\
&&-6\kappa_0+\kappa_0^2) (-64 (1 + 2 \kappa_0)^2  +m^2 (3+12\kappa_0+8\kappa_0^2 - 8\kappa_0^3+\kappa_0^4)-i m (51 + 204 \kappa_0 \nn\\
&&+ 128 \kappa_0^2 - 152 \kappa_0^3 +17 \kappa_0^4) ))S_{lm}(\frac{\pi}{2})+(-8 i\kappa_{S^2} (1 + 2 \kappa_0) (-3 -6 \kappa_0 + \kappa_0^2) (m^2 (-3 \nn\\
&&-6 \kappa_0 + \kappa_0^2)^2 +32 (3 + 12 \kappa_0 + 13 \kappa_0^2 + 2 \kappa_0^3) -16 i m (-3 - 6 \kappa_0 + \kappa_0^2)^2)-96 (1 \nn\\
&&+ 2 \kappa_0) (m (-3 - 6 \kappa_0 + \kappa_0^2)^2 (-1 -2 \kappa_0 + \kappa_0^2) +4 i (1 + 2 \kappa_0)^2 (15 + 30 \kappa_0 \nn\\
&&+ 7 \kappa_0^2)))S'_{lm}(\frac{\pi}{2})-768 (1 + 2 \kappa_0)^3 (-3 - 6 \kappa_0 + \kappa_0^2)S''_{lm}(\frac{\pi}{2})];\nn\\
\mbox{}_{+2}t^{(0)}_2\hspace{-6pt}&=&\hspace{-6pt}-\frac{(1+\kappa_0)^2}{2\lambda_0}S_{lm}(\frac{\pi}{2});\nn\\
\mbox{}_{+2}t^{(1)}_2\hspace{-6pt}&=&\hspace{-6pt}-\frac{i}{4\lambda_0^3}[(m (-3 + (-6 + \kappa_0) \kappa_0)^2 -i (-15 + \kappa_0 (-60 + \kappa_0 (-42 \nn\\
&&+ \kappa_0 (36 + \kappa_0)))))S_{lm}(\frac{\pi}{2})-2 (-3 + (-6 + \kappa_0) \kappa_0) (5 + \kappa_0 (10 + \kappa_0))S'_{lm}(\frac{\pi}{2})];\nn\\
\mbox{}_{+2}t^{(2)}_2\hspace{-6pt}&=&\hspace{-6pt}\frac{1}{16\lambda_0^5(1+2\kappa_0)}[(\kappa_{S^2}(-3 - 6 \kappa_0 + \kappa_0^2) (-m^2 (1 + 2 \kappa_0) (-3 -6 \kappa_0 + \kappa_0^2)^2 -48 (1 \nn\\
&&+ 2 \kappa_0)^2 (-5 - 10 \kappa_0 + 3 \kappa_0^2) +2 i m (-3 - 6 \kappa_0 + \kappa_0^2)^2 (1 + 2 \kappa_0 +3 \kappa_0^2))+6 i (1 \nn\\
&&+ \kappa_0)^2 (1 +2 \kappa_0) (m (-3 - 6 \kappa_0 + \kappa_0^2)^2 +i (3 + 12 \kappa_0 + 32 \kappa_0^2 +40 \kappa_0^3 +\kappa_0^4)))S_{lm}(\frac{\pi}{2})\nn\\
&&+(\kappa_{S^2}(-4 (-3 - 6 \kappa_0 + \kappa_0^2) (-3 - 12 \kappa_0 -11 \kappa_0^2 +2 \kappa_0^3) (m (-3 - 6 \kappa_0 + \kappa_0^2) -4 i (3 \nn\\
&&+ 6 \kappa_0 + \kappa_0^2))) -12 i (1 + \kappa_0)^2 (-11 - 22 \kappa_0 + \kappa_0^2) (-3 -12 \kappa_0 - 11 \kappa_0^2 + 2 \kappa_0^3))S'_{lm}(\frac{\pi}{2})\nn\\
&&+16\kappa_{S^2}((3 + 12 \kappa_0 + 11 \kappa_0^2 - 2 \kappa_0^3)^2) S''_{lm}(\frac{\pi}{2})];\nn\\
\mbox{}_{+2}t^{(1)}_3\hspace{-6pt}&=&\hspace{-6pt}-\frac{(1+\kappa_0)(1+2\kappa_0)}{\lambda_0}S_{lm}(\frac{\pi}{2});\nn\\
\mbox{}_{+2}t^{(2)}_3\hspace{-6pt}&=&\hspace{-6pt}-\frac{1+\kappa_0}{8\lambda_0^3}[(\kappa_{S^2}(-3 - 6 \kappa_0 + \kappa_0^2) (i m (-3 -6 \kappa_0 + \kappa_0^2) + 4 (5 + 10 \kappa_0 + \kappa_0^2))\nn\\
&&+48 (1 + 2 \kappa_0)^2)S_{lm}(\frac{\pi}{2})+8 i\kappa_{S^2} (3 + 12 \kappa_0 + 11 \kappa_0^2 - 2 \kappa_0^3)S'_{lm}(\frac{\pi}{2})] q^3;\nn\\
\mbox{}_{+2}t^{(2)}_4\hspace{-6pt}&=&\hspace{-6pt}-\frac{(1+\kappa_0)^2(1+2\kappa_0)\kappa_{S^2}}{4\lambda_0}S_{lm}(\frac{\pi}{2}).\nn
\eea

\section{Radial equation with delta function source}

Let us consider the second order differential equation
\be
A(r)(B(r)R(r)')'-V(r)R(r)=T(r)\label{deom}
\ee
sourced by a sum of derivatives of delta functions, $T(r)=\sum_{i=0}^{N+2}a_{i}\delta^{(i)}(r-r_0)$. In the main text, $N$ is related to the maximal multipole order considered: $N=0,1,2$ is related to the truncation to monopole (mass), dipole (spin) and quadrupole, respectively. We denote as $R_1(r) $, $R_2(r) $ two independent solutions of the homogeneous equation. The Wronskian 
\bea
W=B(R_1R_2'-R_2R_1')
\eea
is a constant. A general solution is given by 
 \be
 R(r)=X_1 R_1(r)\Theta(r_0-r)+X_2R_2(r)\Theta(r-r_0)+Y_1 R_1(r)+Y_2 R_2(r)+\sum_{i=0}^{N}\beta_i\delta^{(i)}(r-r_0).\label{dif}
 \ee
 The solution can be found by substituting (\ref{dif}) into (\ref{deom}) and solving the linear equations. The coefficients $X_1,X_2$ and $\beta_k$ are unique. The coefficients $Y_1$ and $Y_2$ are free and imposed by fixing the boundary conditions. In this paper, we just need the result up to $N=2$ and we will set $Y_2 = 0$. The solution is 
 \bea
 X_1&=&\frac{-1}{A^5B^3W}[a_0(-A^4B^3R_2)+a_1A^3B^3(AR_2'-A'R_2)+a_2A^2B^2(R_2(-AV-2BA'^2\nn\\
 &&+ABA'')+R_2'(2ABA'+A^2B'))+a_3AB(R_2(-4ABVA'-6A'^3B^2-2A^2B'V\nn\\
 &&+A^2BV'+6AA'A''B^2-A^2B^2A^{(3)})+R_2'(A^2BV+6AB^2A'^2+3A^2BA'B'+2A^3B'^2\nn\\
 &&-3A^2B^2A''-A^3BB''))+a_4(R_2(-A^2BV^2-18AB^2A'^2V-24A'^4B^3-11A^2BA'B'V\nn\\&&-6A^3B'^2V+6A^2B^2A'V'+3A^3BB'V'+7A^2B^2A''V+36AB^3A'^2A''-6A^2B^3A''^2\nn\\
 &&+3A^3BB''V-A^3B^2V''-8A^2B^3A'A^{(3)}+A^3B^3A^{(4)})+R_2'(6A^2B^2A'V+24AB^3A'^3\nn\\&&+4A^3BB'V+12A^2B^2A'^2B'+8A^3BA'B'^2+6A^4B'^3-2A^3B^2V'-24A^2B^3A'A''\nn\\&&-6A^3B^2B'A''-4A^3B^2A'B''-6A^4BB'B''+4A^3B^3A^{(3)}+A^4B^2B^{(3)}))],\\
 X_2&=&X_1(R_2\leftrightarrow R_1,\text{ keeping }W\text{ unflipped}),\label{X2coef}\\
 \beta_0&=&\frac{1}{A^3B^3}(a_2A^2B^2+a_3AB(2AB'+3A'B)+a_4(ABV+AB(8A'B'-6A''B)\nn\\&&+6A^2B'^2-3A^2BB''+12A'^2B^2))),\\
 \beta_1&=&\frac{1}{A^2B^2}(a_3 AB+(3AB'+4A'B)a_4),\\
 \beta_2&=&\frac{a_4}{AB}.
 \eea
The result for $N=1$ can be obtained by substituting $a_4=0$ in the previous solution. Explicitly, 
\bea
X_1&=&\frac{-1}{A^4B^2W}[-A^3B^2R_2a_0+A^2B^2(AR_2'-A'R_2)a_1\nn\\&&+AB((ABA''-2BA'^2-AV)R_2+(A^2B'+2ABA')R_2')a_2\nn\\&&+(R_2(-4ABVA'-6B^2A'^3-2A^2B'V+A^2BV'+6AB^2A'A''-A^2B^2A''')\nn\\&&+R_2'(A^2BV+6AA'^2B^2+3A^2BA'B'+2A^3B'^2-3A^2B^2A''-A^3BB''))a_3],\nn\\
X_2&=&X_1(R_2\leftrightarrow R_1,\text{ keeping }W\text{ unflipped}),\\
\beta_0&=&\frac{a_2 AB+(2AB'+3A'B)a_3}{A^2B^2},\nn\\
\beta_1&=&\frac{a_3}{AB}.\nn
\eea
The result for $N=0$ can be obtained by substituting $a_3=0$ in the previous solution. Explicitly,  
\bea
X_1&=&\frac{-1}{A^3B W}[a_0(-A^2BR_2)+a_1(-ABA'R_2+A^2BR_2')+a_2(R_2(-AV-2A'^2B+ABA'')\nn\\&&+R_2'(2ABA'+A^2B'))],\\
X_2&=&X_1(R_2\leftrightarrow R_1,\text{ keeping }W\text{ unflipped}),\\
\beta_0&=&\frac{a_2}{AB}.
\eea 
All the functions are evaluated at $r=r_0$. 

For near-NHEK region, 
\bea
A(r)&=& (r(r+2\kappa))^{-s},\ B(r)=(r(r+2\kappa))^{s+1}, \label{ABn} \\
V(r)&=&-\frac{3}{4}m^2-s(s+1)+\mathcal E_{lm}-2i s m+\frac{(mr+\kappa n)(\kappa (2si-n)+r(2si-m))}{r(r+2\kappa)},\nn
\eea
with $n=m+\omega/\kappa$ and the two independent solutions are $R_1 = \mathcal R_{lm\omega}^{\text{in}}$, $R_2 = \mathcal R_{lm\omega}^{\text{D}}$ given in \eqref{Rnnn}. 

For the NHEK region, 
\bea
A(r)&=&r^{-2s},\ B(r)=r^{2(s+1)},\nn\\
V(r)&=&-\frac{7}{4}m^2+\mathcal E_{lm}-s(s+1)-\frac{2\Omega (m-is)}{R}-\frac{\Omega^2}{R^2},
\eea
and the two independent solutions are $R_1 =\mathcal W^{\text{in}}_{lm\Omega} $, $R_2 = \mathcal M^{\text{D}}_{lm\Omega}$ given in \eqref{Rnn}.

%\bibliography{refs-6}

\providecommand{\href}[2]{#2}\begingroup\raggedright\endgroup

\end{document}